\documentclass[twocolumn,aps,pra,amsmath]{revtex4-1}

\usepackage{graphicx}

\newcommand{\bfr}{{\bf r}}
\newcommand{\bfx}{{\bf x}}
\newcommand{\bfj}{{\bf j}}
\newcommand{\bfk}{{\bf k}}
\newcommand{\bfR}{{\bf R}}
\newcommand{\bfA}{{\bf A}}
\newcommand{\ua}{\uparrow}
\newcommand{\da}{\downarrow}

\newcommand{\bfq}{{\bf q}}
\newcommand{\bfG}{{\bf G}}

\newcommand{\mbbr}{\underline{\underline{\rm r}}}
\newcommand{\mbbR}{\underline{\underline{\rm R}}}
\newcommand{\mbbPsi}{\Psi}
\newcommand{\gs}{_{\rm gs}}

\newcommand{\vect}[1]{\mathbf{#1}}
\newcommand{\parref}[1]{(\ref{#1})}

\begin{document}

\title{A brief compendium of time-dependent density-functional theory}

\author{Carsten A. Ullrich and Zeng-hui Yang}
\affiliation{Department of Physics and Astronomy, University of Missouri, Columbia, MO 65211, USA}

\date{\today}

\begin{abstract}
Time-dependent density-functional theory (TDDFT) is a formally exact approach to the time-dependent electronic
many-body problem which is widely used for calculating excitation energies. We present a survey of the fundamental framework,
practical aspects, and applications of TDDFT.
This paper is mainly intended for non-experts (students or researchers in other areas) who would
like to learn about the present state of TDDFT without going too deeply into formal details.
\end{abstract}

\maketitle

\section{Preface} \label{sec:I}

This paper presents an introduction to and a survey of time-dependent density-functional theory (TDDFT).
The purpose of the paper is to explain in a nutshell what TDDFT is and what it can do. We will discuss
the basics of the formal framework of TDDFT as well as the current state of the art, skipping over details of the proofs,
and highlight some of the most important applications.
Readers who would like a more detailed treatment and more literature references are encouraged to consult
recent books \cite{Ullrich2012,Marques2012} and review articles \cite{Marques2004,Burke2005b,Casida2012}.

TDDFT is a theoretical approach to the dynamical quantum many-body problem; it can
be used to describe quantum systems that are not stationary. As a consequence, TDDFT provides formally exact and
practically convenient methods to calculate electronic excitation energies.
By contrast, density-functional theory (DFT) is a ground-state theory: in other words,
it is used to find the ground state of a quantum system
and calculate related quantities of interest, such as the ground-state energy.
In many, if not most, situations of practical interest, we have to determine the ground state of the system
before we can study its dynamics or calculate its excitations.

The beginnings of ground-state DFT date back to the years 1964/65 when the famous papers by Hohenberg and Kohn
\cite{Hohenberg1964} and Kohn and Sham \cite{Kohn1965} were published. Since then, DFT has developed into a
dominating method for electronic structure calculations in physics, chemistry, materials science, and many other areas
(see Ref. \cite{Burke2012} for a recent up-to-date account of DFT). Although TDDFT is of much more recent origin \cite{Runge1984},
it now has reached a similar status for calculating electronic excitations.

TDDFT uses many familiar concepts from DFT, most prominently, the Kohn-Sham idea of replacing the real interacting many-body
system by a noninteracting system that reproduces the same density.
But there are also many concepts that are unique to the time-dependent case, such as memory and initial-state dependence.
To gain a thorough understanding of TDDFT it is hence advisable to begin with a study of the basic concepts of DFT.
We refer the reader to the very nice introductions to DFT by Capelle \cite{Capelle2006} and by Burke and Wagner \cite{Burke2013}.
There exist a number of books on DFT, some of which
are very accessible to newcomers in the field \cite{Koch,Sholl2009}, others are more advanced \cite{Engel2011}.

\section{Ground-state DFT in a nutshell} \label{sec:II}

\subsection{The many-body problem} \label{subsec:II.A}

We consider a system of $N$ interacting electrons that is described by  the nonrelativistic Schr\"odinger equation:
\begin{equation} \label{II.A.1}
\hat H_0 \Psi_j(\bfr_1,\ldots,\bfr_N) = E_j \Psi_j(\bfr_1,\ldots,\bfr_N) \:,\quad j=0,1,2,\ldots \:.
\end{equation}
For a given $D$-dimensional system,
the $j$th eigenstate $\Psi_j(\bfr_1,\ldots,\bfr_N)$ is a function of $DN$ spatial variables.
In the following, we  use the abbreviation $\Psi_j$. Of particular interest to us is the ground-state wave function $\Psi\gs$.

The many-body Hamiltonian is given by
\begin{equation} \label{II.A.2}
\hat H = \hat T + \hat V + \hat W \:,
\end{equation}
where the kinetic-energy and scalar potential operators are
\begin{equation} \label{II.A.3}
\hat T = \sum_{j=1}^N -\frac{\nabla_j^2}{2} \:, \qquad \quad
\hat V = \sum_{j=1}^N v(\bfr_j) \:,
\end{equation}
and the electron-electron interaction operator is
\begin{equation} \label{II.A.5}
\hat W = \sum_{i,j=1 \atop i\ne j}^N w(|\bfr_i - \bfr_j|) \:.
\end{equation}
Notice that we use atomic (Hartree) units throughout, i.e., $m=e=\hbar=1$.
The electron-electron interaction is usually taken to be the Coulomb interaction, $w(|\bfr - \bfr'|)= 1/|\bfr - \bfr'|$,
but other forms of two-particle interactions, or zero interactions, are also possible.

The single-particle potential $v(\bfr)$ describes the total potential acting on the electrons.
If one is interested in describing the properties of matter (atoms, molecules, or solids), $v(\bfr)$ is the
sum of the Coulomb potentials of the atomic nuclei.
However, to define the formal framework of DFT it is not necessary to specify where the potential comes from,
as long as it has a mathematically well-behaved form.

From the solutions of the Schr\"odinger equation we calculate the expectation value of an observable in the $j$th eigenstate:
\begin{equation} \label{II.A.6}
O_j = \langle \Psi_j | \hat{O} | \Psi_j\rangle .
\end{equation}
Here, $\hat{O}$ is a Hermitian operator corresponding to a quantum mechanical observable.

Let us make two remarks on our formulation of the many-body problem.

(a) We implicitly made the Born-Oppenheimer approximation (see Section \ref{subsec:VII.C}).
In other words, if our system contains nuclear degrees of freedom (as is the case in all forms of real
matter), we treat them classically. The many-body wave functions
therefore depend only on the electronic coordinates $(\bfr_1,\ldots,\bfr_N)$, and the nuclei only act as sources of
scalar potentials. In Section \ref{sec:VIII} we will briefly discuss what happens if this approximation is not made
and the electronic and nuclear degrees of freedom are coupled.

(b) We have not indicated any spin indices, which was done mainly for notational simplicity.
In other words, $\Psi_j(\bfr_1,\ldots,\bfr_N)$ describes spinless electrons.
Including spin, the many-body wave function can be written as $\Psi(\bfx_1,\ldots,\bfx_N)$,
where $\bfx_i = (\bfr_i,\sigma_i)$ denotes the spatial and spin coordinate of the $i$th electron.

\subsection{The basic idea behind DFT} \label{subsec:II.B}

Everything we wish to know about our system  (energy, geometry, excitation spectrum, etc.)
can be obtained from the wave functions, see Eq. (\ref{II.A.6}). The exact wave functions can be
calculated if the system is small, with no
more than one or two electrons, but this becomes very difficult if $N$ is greater
than that: the many-body Schr\"odinger equation becomes too hard to solve, and the usefulness of the wave function
itself becomes more and more questionable for large $N$ \cite{Kohn1999}.

There exist many approaches to find approximate solutions of the many-body problem. So-called ``wave-function based techniques''
such as Hartree-Fock (HF) or configuration-interaction (CI) attempt to find variational solutions of the
Schr\"odinger equation using expansions of the wave function in terms of Slater determinants. This approach
has been very successful in theoretical chemistry, but has its limitations for large systems.

{\em The essence of the density-functional approach is that it is in principle possible to obtain all desired
information about an $N$-electron system without having to calculate its full wave function: instead, all one needs
is the one-particle probability density of the ground state,}
\begin{equation} \label{II.1}
n_0(\bfr) = N\int d^3r_2\ldots \int d^3r_N |\Psi\gs(\bfr,\bfr_2,\ldots,\bfr_N)|^2 \:.
\end{equation}

\begin{figure}
\centering
\includegraphics[width=\linewidth]{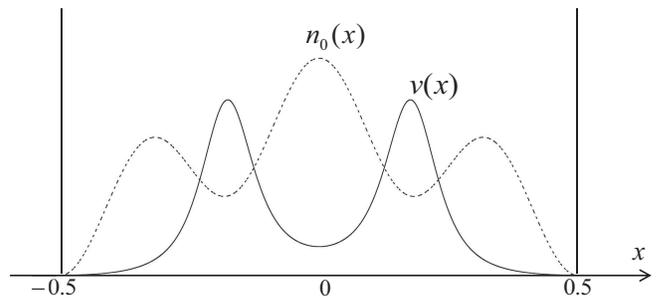}
\caption{The density $n_0(x)=\cos^2(\pi x) + \cos^2(3\pi x)$ (dashed line, scaled by a factor 60), for $-\frac{1}{2}<x<\frac{1}{2}$,
is the ground-state density of the potential $v(x)$ (full line) which was constructed using Eq. (\ref{II.3}).} \label{fig1}
\end{figure}

This can be mathematically proven (see below), but before doing so it is helpful to give a simple illustration.
Consider a one-electron system which satisfies the Schr\"odinger equation
\begin{equation} \label{II.2}
\left[-\frac{\nabla^2}{2} + v(\bfr)\right] \varphi_j(\bfr) = \varepsilon_j \varphi_j(\bfr) \:.
\end{equation}
The usual procedure is to solve this equation for a given potential $v(\bfr)$ and determine the
ground-state probability density as $n_0(\bfr) = |\varphi_0(\bfr)|^2$.
But now imagine the reverse situation: we are given a density function $n_0(\bfr)$, normalized to 1,
and we ask in what potential this is the ground-state density. Assuming that the wave functions are real
so $\varphi_0(\bfr)=\sqrt{n_0(\bfr)}$,
the Schr\"odinger equation (\ref{II.2}) is easily inverted, and we obtain
\begin{equation} \label{II.3}
v(\bfr) = \frac{\nabla^2 n_0(\bfr)}{4n_0(\bfr)} - \frac{|\nabla n_0(\bfr)|^2}{8n_0(\bfr)^2} \:.
\end{equation}
A one-dimensional example is given in Fig. \ref{fig1}.

What has been accomplished? From the ground-state density $n_0(\bfr)$ we were able to reconstruct the potential
$v(\bfr)$. But this means that we have reconstructed the Hamiltonian $\hat H$ of the system, and we can thus solve
the Schr\"odinger equation  (\ref{II.2}) and get all the wave functions!
This logical chain can be represented as follows:
\begin{equation} \label{II.4}
n_0(\bfr) \to v(\bfr) \to \hat H \to \{\Psi_j\}.
\end{equation}
The reconstruction of the potential from the density is easy for one-electron systems. For interacting systems with
many electrons there is no explicit formula such as Eq. (\ref{II.3}). Nevertheless, there exists a {\em unique}
potential for each mathematically well-behaved density function such that it is the ground-state density in this potential.
This was proved by Hohenberg and Kohn in 1964 \cite{Hohenberg1964}.

The {\em Hohenberg-Kohn theorem} states that it is impossible for two different potentials, $v(\bfr)$ and $v'(\bfr)$, to
produce the same ground-state density ($v'$ is considered to be different from $v$ if it is not just $v$ shifted by a constant).
In other words, the relationship between potentials and ground-state densities is one-to-one:
\begin{equation} \label{II.5}
n_0(\bfr) \leftrightarrow v(\bfr) \:.
\end{equation}
The proof of this theorem is relatively straightforward, making use of the Rayleigh-Ritz minimum principle.
It can be found in any textbook on DFT, so we won't repeat it here.

Formally, this logical dependence of the wave functions on the ground-state density constitutes a {\em functional} relationship,
which is written as $\Psi_j[n_0]$. Hence, the name density-functional theory. Every quantum mechanical observable thus
can be written as a density functional.

In particular, the total energy functional of a system with potential $v_0(\bfr)$ is
\begin{equation} \label {II.6}
E_{v_0}[n]=\langle \Psi[n]|\hat T + \hat V_0 + \hat W | \Psi[n]\rangle \:,
\end{equation}
where $n$ is some $N$-electron density and $\Psi[n]$ is that ground-state wave function which reproduces this density.
The energy functional (\ref{II.5}) is minimized by the ground-state density $n_0$ which belongs to $v_0$,
and then becomes equal to the ground-state energy:
\begin{eqnarray}
E_{v_0}[n] > E_0 \quad \mbox{for} \quad n(\bfr) \ne n_0(\bfr)\:, \nonumber\\
E_{v_0}[n] = E_0 \quad \mbox{for} \quad n(\bfr) = n_0(\bfr) \:.
\end{eqnarray}

\subsection{The Kohn-Sham approach} \label{subsec:II.C}

The fact that all observables are functionals of the density opens up the way for an
enormous computational simplification, since the density is a function of only $D$ variables
(and not of $DN$ variables as the wave function). But how can we take advantage of this in practice?
To obtain the density one still needs to solve the full many-body problem.
This means that nothing
has been gained, unless we find a way to bypass the full Schr\"odinger equation and obtain the density
in some other, easier way, at least approximately. Fortunately, a very elegant method exists to do this, known as the {\em Kohn-Sham
formalism} \cite{Kohn1965}.

We use the following trick: we define a {\em noninteracting} system in such a way that it reproduces the exact
ground-state density of the {\em interacting} system. This means that we can calculate the exact density as the
sum of squares of single-particle orbitals,
\begin{equation} \label{II.10}
n_0(\bfr) = \sum_{j=1}^N |\varphi_j(\bfr)|^2\:,
\end{equation}
where the orbitals satisfy the following equation;
\begin{equation} \label{II.11}
\left[ -\frac{\nabla^2}{2} + v_s[n](\bfr)\right] \varphi_j(\bfr) = \varepsilon_j \varphi_j(\bfr) \:.
\end{equation}
This equation is known as {\em Kohn-Sham equation}; it is formally a single-particle Schr\"odinger equation,
like Eq. (\ref{II.2}). However, the potential $v_s$ is very special: it is {\em defined} to be that single-particle
potential that produces orbitals which give the {\em exact} ground-state density of the interacting system via Eq. (\ref{II.10}).
It is therefore a functional of the density, $v_s[n](\bfr)$.

The trick is now to write the unknown effective potential $v_s[n]$ in a smart way. No doubt, the given external potential $v_0(\bfr)$
will make a contribution to it. The remainder, $v_s[n]-v_0(\bfr)$, then accounts for the electronic many-body effects.
A large portion of the latter is made by the classical Coulomb potential associated with a given density
distribution, also known as the Hartree potential,
\begin{equation} \label{II.12}
v_{\rm H}(\bfr) = \int d^3r' \frac{n(\bfr')}{|\bfr - \bfr'|} \:.
\end{equation}
And whatever is left is called the {\em exchange-correlation} (xc) potential, $v_{\rm xc}[n](\bfr)$, so that
\begin{equation}
v_s[n](\bfr) = v_0(\bfr) + v_{\rm H}(\bfr) + v_{\rm xc}[n](\bfr) \:.
\end{equation}
It turns out that the solution of the Kohn-Sham equation (\ref{II.11})---that is, the density (\ref{II.10})--- is precisely that
density which minimizes the total energy functional (\ref{II.6}). The connection is made by rewriting the energy as follows:
\begin{eqnarray}
E_{v_0}[n] &=& T[n] + \int d^3r \: v_0(\bfr) n(\bfr) + W[n] \nonumber\\
&=&
T_s[n] + \!\int \! d^3r  v_0(\bfr) n(\bfr) + (T[n] - T_s[n] + W[n] ) \nonumber\\
&\equiv&
T_s[n] + \int \! d^3r v_0(\bfr) n(\bfr) + E_{\rm H}[n] + E_{\rm xc}[n] \:. \label{II.13}
\end{eqnarray}
Here, $T[n]$ is the kinetic-energy functional of an {\em interacting} system, whereas $T_s[n]$ is the
kinetic-energy functional of a {\em noninteracting} system. Neither $T[n]$ nor $T_s[n]$ are
known as explicit density functionals, but it is very straightforward to write down $T_s[n]$ as
an explicit functional of the orbitals:
\begin{equation} \label{II.14}
T_s[n] =-\frac{1}{2} \sum_{j=1}^N \varphi_j^*(\bfr) \nabla^2 \varphi_j(\bfr) \;,
\end{equation}
where the orbitals ${\varphi_j(\bfr)}$ come from the Kohn-Sham equation (\ref{II.11}) are are hence implicit density functionals.

In the last line of Eq. (\ref{II.13}) we define the Hartree energy
\begin{equation}\label{II.15}
E_{\rm H}[n] = \frac{1}{2} \int d^3r \int d^3r' \: \frac{n(\bfr) n(\bfr')}{|\bfr - \bfr'|}
\end{equation}
and the xc energy as
\begin{equation} \label{II.16}
E_{\rm xc}[n] = T[n] - T_s[n] + W[n] - E_{\rm H}[n] \:.
\end{equation}
This shows that the xc potential is given by the following functional derivative:
\begin{equation} \label{II.17}
v_{\rm xc}(\bfr) = \frac{\delta E_{\rm xc}[n]}{\delta n(\bfr)} \:.
\end{equation}
It is straightforward to show that the total energy (\ref{II.13}) can be expressed as follows:
\begin{eqnarray} \label{II.18}
E_{v_0}[n] &=& \sum_{j=1}^N \varepsilon_j - E_{\rm H}[n] -\int d^3r \: v_{\rm xc}(\bfr) n(\bfr) +  E_{\rm xc}[n] \:.
\nonumber\\
&&
\end{eqnarray}

\subsection{Discussion and exact properties} \label{subsec:II.D}

Let us now summarize some of the most important properties of the Kohn-Sham approach. Our discussion is by no
means complete, but the following properties will be relevant for the time-dependent case as well.

{\em Meaning of the wave function.} The Kohn-Sham system is noninteracting, so its total $N$-particle wave function can be written as a
single Slater determinant:
\begin{equation}
\Psi^{\rm KS}\gs(\bfr_1,\ldots,\bfr_N) = \frac{1}{\sqrt{N!}} \mbox{det}\{\varphi_j\} \:.
\end{equation}
The Kohn-Sham Slater determinant has only one purpose: to reproduce the exact ground-state density when substituted in Eq. (\ref{II.1}).
It is {\em not} meant to reproduce the exact ground-state wave function, i.e., $\Psi\gs^{\rm KS} \ne \Psi\gs$ in general.

{\em Meaning of the Kohn-Sham energies.}
The energy eigenvalues $\varepsilon_j$ do not have a rigorous physical meaning, except for the highest occupied eigenvalue.
We have
\begin{equation}\label{IN}
\varepsilon_N(N) = E(N)-E(N-1) = -I(N) \:,
\end{equation}
i.e., the highest occupied eigenvalue of the $N$-particle system equals minus the ionization energy of the $N$-particle system,
and
\begin{equation}\label{AN}
\varepsilon_{N+1}(N+1) = E(N+1)-E(N) = -A(N) \:,
\end{equation}
i.e., the highest occupied eigenvalue of the $N+1$-particle system equals minus the electron affinity of the $N$-particle system.

Eigenvalue differences $\varepsilon_a-\varepsilon_i$, where $a$ labels an unoccupied single-particle state and $i$ an occupied one,
should not be interpreted as excitation energies of the many-body system (although they often are).

{\em Asymptotic behavior of the Kohn-Sham potential.}
A neutral atom with $N$ electrons has the nuclear potential $v_0(\bfr) = -N/r$, and
its Hartree potential behaves as $v_{\rm H}(\bfr) \to N/r$ for $r\to \infty$. If an electron is far away in the outer regions
of the atom, it should see the Coulomb potential of the remaining positive ion. This implies that the xc potential must
behave asymptotically as
\begin{equation} \label{asym}
v_{\rm xc}(\bfr) \to -\frac{1}{r}
\end{equation}
for large $r$, for any finite system.
The asymptotic behavior of $v_{\rm xc}(\bfr)$ reflects the fact that the Kohn-Sham formalism
is free of self-interaction: for a 1-electron system, the Hartree and xc potential cancel exactly.

{\em Spin-dependent formalism.} In practice, the Kohn-Sham formalism is usually written down and applied in its
spin-polarized form, even if the system does not have a net spin polarization. We then have
\begin{equation}
\left[ -\frac{\nabla^2}{2} + v_{0\sigma}(\bfr)+v_{\rm H}(\bfr) + v_{\rm xc\sigma}(\bfr)\right] \varphi_{j\sigma}(\bfr)
= \varepsilon_{j\sigma} \varphi_{j\sigma}(\bfr) \:,
\end{equation}
where $\sigma=\uparrow,\downarrow$. Here, the external potential $v_{0\sigma}$ carries a spin index, which could come from
a static magnetic field, and the spin-polarized xc potential is defined as a functional of the spin-up and spin-down density:
\begin{equation}
v_{\rm xc \sigma}[n_\uparrow,n_\downarrow](\bfr) = \frac{\delta E_{\rm xc}[n_\uparrow,n_\downarrow]}{\delta n_\sigma(\bfr)} \:,
\end{equation}
where
\begin{equation}
n_\sigma(\bfr) = \sum_{j=1}^{N_\sigma} |\varphi_{j\sigma}(\bfr)|^2 \:.
\end{equation}

{\em Exact exchange.}
The xc energy can be decomposed into exchange and correlation energy. The exact exchange energy is given by
\begin{eqnarray}
E_{\rm x}^{\rm exact}[n]&=& -\frac{1}{2} \sum_{\sigma = \ua,\da} \sum_{i,j=1}^{N_\sigma}\int d^3r \int d^3r'
\nonumber\\
&&\times
\frac{\varphi_{i\sigma}^*(\bfr)
\varphi_{j\sigma}(\bfr')\varphi_{i\sigma}(\bfr') \varphi_{j\sigma}^*(\bfr)}{|\bfr - \bfr'|} \:,
\end{eqnarray}
where the $\varphi_{j\sigma}(\bfr)$ are the exact Kohn-Sham orbitals. $E_{\rm x}^{\rm exact}$ is a
so-called {\em implicit} density functional.

\subsection{DFT in practice} \label{subsec:II.E}

The number of applications of DFT in various areas of science and engineering is almost impossible to count.
Nice practical introductions from the perspectives of chemistry and of materials science, respectively, can be found
in recent books and review articles \cite{Koch,Sholl2009,Neese2009}.
Here, we will only make some general remarks and discuss a couple of representative examples.

Even though DFT is in principle exact (as we have emphasized), any practical application necessarily involves
two kinds of approximations: (i) the xc energy functional $E_{\rm xc}[n]$, and the xc potential following from it via Eq. (\ref{II.17}),
are not exactly known and need to be approximated; (ii) the Kohn-Sham equation (\ref{II.11}) needs to be solved
using some computational scheme, which can introduce various types of numerical inaccuracies.

Over the years, many approximate xc functionals have been proposed; some of them using physical arguments, constraints
and exact conditions, others using parametrizations combined with fitting to reference data. Which functional should
one choose? This question cannot be easily answered in general \cite{RAppoport2009} but requires some experience.
Practitioners of DFT who use popular software packages of quantum chemistry or solid-state physics often encounter
daunting choices between many different menu options for $v_{\rm xc}$. Some functionals have turned out to be
more popular and successful than others, and are typically chosen in the majority of applications.

The xc energy of any system can be written as
\begin{equation}
E_{\rm xc}[n] = \int d^3r \: e_{\rm xc}[n](\bfr) \:,
\end{equation}
where $e_{\rm xc}[n](\bfr)$ is the xc energy density, whose dependence on the density is, in general, nonlocal:
$e_{\rm xc}$ at a particular point $\bfr$ is determined by the density $n(\bfr')$ at all points in space.
The goal is to approximate $e_{\rm xc}[n](\bfr)$.

Much of the success of DFT can be attributed to the fact
that a very simple approximation, the local-density approximation (LDA), gives very useful results in many circumstances.
The LDA has the following form:
\begin{equation}
E_{\rm xc}^{\rm LDA}[n] = \int d^3r \: e_{\rm xc}^h(n(\bfr))\:.
\end{equation}
Here, the xc energy density of a homogeneous electron liquid,  $e_{\rm xc}^h(\bar n)$ (which
is simply a function of the uniform density $\bar n$), is evaluated at the local density at point $\bfr$ of the actual
inhomogeneous physical system: $e_{\rm xc}^h(n(\bfr)) = \left.e_{\rm xc}^h(\bar n)\right|_{\bar n = n(\bfr)} $.
The so defined $E_{\rm xc}^{\rm LDA}[n]$ is exact in the limit where the system becomes uniform,
and should be accurate when the system varies only slowly in space.

The LDA requires $e_{\rm xc}^h(\bar n)$ as input \cite{GiulianiVignale}. We can write
\begin{equation}
e_{\rm xc}^h(\bar n) = e_{\rm x}^h(\bar n) + e_{\rm c}^h(\bar n) \:,
\end{equation}
where the exchange energy density can be calculated exactly using Hartree-Fock theory; the result
(for the spin-unpolarized electron liquid) is
\begin{equation}
e_{\rm x}^h(\bar n) = -\frac{3}{4}\left(\frac{3}{\pi}\right)^{1/3} \bar n^{4/3} \:.
\end{equation}
This gives the following expression for the LDA exchange potential:
\begin{eqnarray}
v_{\rm x}^{\rm LDA}(\bfr) &=& \frac{\delta}{\delta n(\bfr)} \left[ -\frac{3}{4}\left(\frac{3}{\pi}\right)^{1/3}\int d^3r'
 n(\bfr')^{4/3}\right]
 \nonumber\\
 &=& -\left(\frac{3}{\pi}\right)^{1/3} n(\bfr)^{1/3} \:.
\end{eqnarray}
The correlation energy density $e_{\rm c}^h(\bar n)$ is not exactly known, but very accurate numerical results
exist from quantum Monte Carlo calculations. Based on these results, parametrizations for the correlation
energy of the homogeneous electron liquid have been derived \cite{Vosko1980,Perdew1981,Perdew1992}.

The LDA generally performs very well across the board. It produces atomic and molecular total ground-state energies
within 1-5\% of the exact value, and yields molecular equilibrium distances and geometries within about 3\%.
For solids, Fermi surfaces in metals come out within a few percent, lattice constants of solids within about 2\%,
and vibrational frequencies and phonon energies are obtained within a few percent as well.

On the other hand, the LDA has several shortcomings. For instance, the LDA is not self-interaction free; as a
consequence, the xc potential goes to zero exponentially fast and not as $-1/r$ [Eq. (\ref{asym})].
This causes the Kohn-Sham energy eigenvalues to be too low in magnitude in general; in particular,
the highest occupied eigenvalue $\varepsilon_N$ underestimates the ionization energy typically by 30--50\%.
The LDA does not produce any stable negative ions, and it underestimates the band gap in solids.
Dissociation of heteronuclear molecules in LDA produces ions with fractional charges.

Overall, the LDA often gives good
results in solid-state physics and materials science, but it
is usually not accurate enough for many chemical applications.

\begin{table}[tb]
{
\caption{Mean absolute errors in several molecular properties calculated for various test sets \protect\cite{Staroverov2003}.
\label{table1} }
}
{\begin{tabular}{lrrrr}
\hline
 & & \\[-8pt]
& Formation & Ionization & Equilibrium  & Vibrational \\
& enthalpy$^a$  & potential$^b$  & bond length$^c$  & frequency$^d$  \\[2pt]
\hline
 & & \\[-6pt]
HF   &  211.54 & 1.028 & 0.0249 & 136.2  \\[3pt]
LSDA &  121.85 & 0.232 & 0.0131 &  48.9  \\[3pt]
BLYP &    9.49 & 0.286 & 0.0223 &  55.2  \\[3pt]
PBE  &   22.22 & 0.235 & 0.0159 &  42.0  \\[3pt]
B3LYP&    4.93 & 0.184 & 0.0104 &  33.5  \\[3pt]
\hline\\
\end{tabular}
}

{\small $^a$ For a test set of 223 molecules (in kcal/mol).}

{\small $^b$ For a test set of 223 molecules (in eV), evaluated from the total-energy differences between the cation and the corresponding neutral,
for their respective geometries.}

{\small $^c$ For a test set of 96 diatomic molecules (in \AA).}

{\small $^d$ For a test set of 82 diatomic molecules (in cm$^{-1}$).}

\end{table}

The LDA can be improved by including a dependence not only on the local density itself, but also on gradients of the density.
This defines the so-called generalized gradient approximation (GGA), which has the following generic form:
\begin{equation}
E_{\rm xc}^{\rm GGA}[n_\ua,n_\da] = \int d^3r \: e_{\rm xc} \! \left(n_\ua(\bfr),n_\da(\bfr),
\nabla n_\ua(\bfr),\nabla n_\da(\bfr)\right) .
\end{equation}
There exist hundreds of different GGA functionals, and it is impossible to list all of them here. Among the most famous ones
are the B88 exchange functional \cite{Becke1988}, the LYP correlation functional \cite{Lee1988} (which, combined together, give the
BLYP functional), and the PBE functional \cite{Perdew1996}.
The exchange part of the latter has the following form:
\begin{equation}
E_{\rm x}^{\rm PBE} = \int d^3r e_{\rm x}^h(n)\left[ 1 + \kappa - \frac{\kappa}{1 + \beta \pi^2 s^2/3\kappa} \right],
\end{equation}
where $s(\bfr) = |\nabla n(\bfr)|/2n(\bfr)k_F(\bfr)$, $k_F(\bfr)$ is the local Fermi wavevector,
and  $\kappa$ and $\beta$ are given parameters.

The GGAs have been crucial in the great success story of DFT over the past couple of decades, due to their
accuracy combined with computational simplicity. However, improvements are still desirable. One of the
most important breakthroughs has been the development of the so-called hybrid functionals, which mix in
a fraction of exact exchange:
\begin{equation}
E_{\rm xc}^{\rm hybrid} = a E_{\rm x}^{\rm exact} + (1-a) E_{\rm x}^{\rm GGA} + E_{\rm c}^{\rm GGA} \:,
\end{equation}
where $a$ is a mixing coefficient that has a value of around $0.25$. The most famous hybrid functional
is B3LYP \cite{Stephens1994}, which nowadays has become the workhorse of computational chemistry.
It should be noted that the exact exchange mixed in here prevents the easy construction of a local xc potential,
so hybrid functionals are defined in the so-called generalized-Kohn-Sham scheme \cite{Seidl1996,Kummel2008}.

Table \ref{table1} gives an assessment of various approximate xc functionals, carried out for large molecular
test sets \cite{Staroverov2003}. All xc functionals perform much better than Hartree-Fock. It is evident that the B3LYP
functional gives the best overall results, with accuracies that come close to the requirements for predicting chemical
reactions (the so-called ``chemical accuracy'' of 1 kcal/mol).

In solids, hybrid functionals such as B3LYP perform less well, due to the fact that they do not
reduce to the exact homogeneous electron gas limit \cite{Paier2007}. A detailed assessment of the performance
of modern density functionals for bulk solids was given by Czonka {\em et al.} \cite{Czonka2009}.
Generally speaking, GGA functionals do not improve the lattice constants in nonmolecular solids obtained with LDA (which
are already very good!):
while LDA systematically underestimates lattice constants, GGA overestimates them. Vice versa, bulk moduli and
phonon frequencies are typically overestimated by LDA and underestimated by GGA. This clearly affects many
properties of solids which are volume-dependent such as their magnetic behavior. Some typical results for lattice constants
are given in Table \ref{table2}.

\begin{table}[tb]
{
\caption{Equilibrium lattice constants of some representative bulk solids \protect\cite{Czonka2009}.
The experimental data includes a subtraction of zero-point motion effects.
PBEsol is a variant of the PBE functional \protect\cite{Perdew2008}, and TPSS is a
so-called meta-GGA functional, which contains a dependence on density gradients and on the kinetic-energy density \protect\cite{Tao2003}.
\label{table2} }
}
{\begin{tabular}{cccccc}
\hline
 & & \\[-8pt]
& LDA  & PBE  & PBEsol  & TPSS & Experiment  \\[2pt]
\hline
 & & \\[-6pt]
Li   &  3.363 & 3.429 & 3.428 & 3.445 & 3.449  \\[3pt]
Na   &  4.054 & 4.203 & 4.167 & 4.240 & 4.210  \\[3pt]
Cu   &  3.517 & 3.628 & 3.562 & 3.575 & 3.595  \\[3pt]
Si   &  5.403 & 5.466 & 5.431 & 5.451 & 5.416 \\[3pt]
NaCl &  5.465 & 5.700 & 5.602 & 5.703 & 5.565  \\[3pt]
MgO  &  4.168 & 4.255 & 4.223 & 4.237 & 4.184 \\[3pt]
\hline\\
\end{tabular}
}

\end{table}

A particular class of hybrid functionals, called {\em range-separated hybrids}, has attracted much interest lately
\cite{Baer2010}. The basic idea is to separate the Coulomb interaction into a short-range (SR) and a long-range (LR) part:
\begin{equation} \label{2.fsep}
\frac{1}{|\bfr - \bfr'|} = \frac{f(\mu|\bfr - \bfr'|)}{|\bfr - \bfr'|} + \frac{1-f(\mu|\bfr - \bfr'|)}{|\bfr - \bfr'|} \:,
\end{equation}
where the function $f$ has the properties $f(\mu x\to 0)=1$ and $f(\mu x\to \infty)=0$. Common examples are
$f(\mu x) = e^{-\mu x}$ and $f(\mu x) = \mbox{erfc}(\mu x)$.
The separation parameter $\mu$ is determined either empirically \cite{Iikura2001,Yanai2004,Baer2005,Gerber2005,Vydrov2006} or using physical
arguments \cite{Livshits2007,Baer2010}.
The resulting range-separated hybrid xc functional then has the following generic form:
\begin{equation}
E_{\rm xc} = E_{\rm x}^{\rm SR-DFA} + E_{\rm x}^{\rm LR-HF} + E_{\rm c}^{\rm DFA} \:,
\end{equation}
where DFA stands for any standard density-functional approximation such as the LDA or GGA.
The main strength of range-separated hybrids is that they have the correct (Hartree-Fock)
long-range asymptotic behavior, and at the same time take advantage of the good short-range behavior of LDA or GGA.
This, in turn, leads to a significant improvement in properties such as the polarizabilities of long-chain molecules,
bond dissociation, and, particularly importantly for TDDFT, Rydberg and charge-transfer excitations (see Section \ref{subsec:VI.C}).

This concludes our very brief survey of ground-state DFT. Let us now come to the dynamical case.


\section{Survey of dynamical phenomena} \label{sec:III}

The stationary many-body problem was defined in Section \ref{subsec:II.A}. Solving the Schr\"odinger equation
(\ref{II.A.1}) allows us to obtain the eigenstates of an $N$-particle system. The time-dependent Schr\"odinger equation
is given by
\begin{equation} \label{III.1}
i \frac{\partial}{\partial t} \Psi_j(\bfr_1,\ldots,\bfr_N,t) = \hat H(t) \Psi_j(\bfr_1,\ldots,\bfr_N,t) \:,
\end{equation}
where the time-dependent Hamiltonian is defined as
\begin{equation}\label{III.2}
\hat H(t) = \hat T + \hat V(t) + \hat W \:.
\end{equation}
The time-dependent Hamiltonian has the same kinetic-energy and electron-electron interaction parts $\hat T$ and $\hat W$
as the static Hamiltonian (\ref{II.A.2}), but it features an external potential operator that is explicitly time-dependent:
\begin{equation} \label{III.3}
\hat V(t) = \sum_{j=1}^N v(\bfr_j,t) \:.
\end{equation}

The time-dependent Schr\"odinger equation (\ref{III.1}) formally represents an initial-value problem.
We define a time $t_0$ as our initial time (often, $t_0=0$), and we
start with a given initial many-body wave function of the system, $\Psi(t_0) \equiv \Psi_0$ (notice that this is
not necessarily the ground state). This state is then propagated forward in time, describing how the system evolves under the
influence of the time-dependent potential $v(\bfr,t)$.
In many situations of practical interest, the time-dependent single-particle potential can be written as
\begin{equation} \label{III.4}
v(\bfr,t) = v_0(\bfr) + \theta (t-t_0)v_1(\bfr,t) \:,
\end{equation}
i.e., the potential is static and equal to $v_0$ until time $t_0$ when an explicitly time-dependent additional potential
$v_1(t)$ is switched on.

The time-dependent wave function
allows us to calculate whatever observable we may be interested in,
\begin{equation} \label{III.5}
O(t) = \langle \Psi(t) | \hat{O} | \Psi(t)\rangle .
\end{equation}
Here, $O(t)$ is the time-dependent expectation value of the Hermitian operator
$\hat{O}$ corresponding to a quantum mechanical observable.
Two key quantities for TDDFT are the time-dependent density and current density, $n(\bfr,t)$ and $\bfj(\bfr,t)$.
They can be defined via the one-particle density operator and current density operator,
\begin{eqnarray} \label{III.6}
\hat{n}(\bfr) &=& \sum_{i=1}^N \delta(\bfr - \bfr_i) \\
\hat{\bfj}(\bfr) &=& \sum_{i=1}^N[\nabla_i\delta(\bfr - \bfr_i) + \delta(\bfr - \bfr_i)\nabla_i]
\end{eqnarray}
so that $n(\bfr,t) = \langle \Psi(t) | \hat{n}(\bfr) |\Psi(t)\rangle$ and similar for $\bfj(\bfr,t)$.
A connection between density and current density is provided by the continuity equation,
\begin{equation} \label{III.7}
i \frac{\partial}{\partial t} n(\bfr,t) = -\nabla \cdot \bfj(\bfr,t)\:.
\end{equation}

There are many different types of quantum mechanical time evolution that are of practical interest.
Many of them belong to one of the following two generic scenarios.

\begin{figure}
\centering
\includegraphics[width=0.6\linewidth]{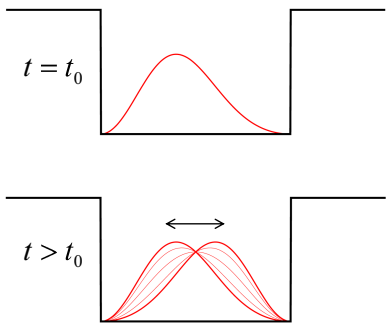}
\caption{First scenario of time evolution: the external potential is static, but the system starts with a nonequilibrium
initial state. The density then oscillates back and forth.} \label{fig2}
\end{figure}

{\em First scenario.} Consider a system that starts from a nonequilibrium initial state, and then
freely evolves in a static potential. A simple one-dimensional example is illustrated in Fig. \ref{fig2}:
at the initial time $t_0$, the density has an asymmetric shape which clearly does not come from an eigenstate of the square-well potential.
The density is then ``released'' and starts to oscillate back and forth, while the
square well potential remains static \footnote{Strictly speaking, the wave function should penetrate a little
bit into the barrier if the square well has finite depth. We ignore this for simplicity and clarity of presentation.}.

This kind of free time evolution occurs in practice when the system is subject to a sudden switching or a
short ``kick'' at the initial time, and is then left to itself. For example, charge-density oscillations
that are triggered in this way play an important role in the field of ``plasmonics'' \cite{Schuller2010}.

\begin{figure}
\centering
\includegraphics[width=0.6\linewidth]{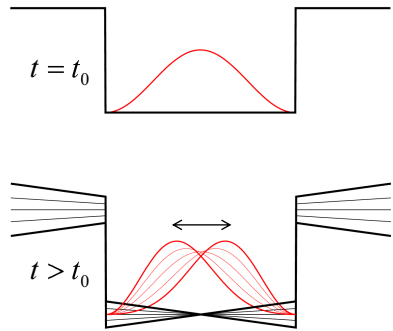}
\caption{Second scenario of time evolution: the system starts from the ground state and evolves under a time-dependent
external potential that is switched on at the initial time $t_0$.} \label{fig3}
\end{figure}

{\em Second scenario.} Consider now a system that is initially in the ground state, and is then subject
to a time-dependent potential that is switched on at time $t_0$. This is illustrated in Fig. \ref{fig3} for
a square well potential that is ``shaken'' by superimposing it with a time-dependent linear potential,
which again leads to an oscillating density.

For example, this scenario takes place if an atom or molecule is hit by a strong laser pulse: the wave function
is driven by the external field and gets ``shaken up'', which can then lead to ionization.

TDDFT will allow us to describe both dynamical scenarios formally exactly for arbitrary many-body systems.
To do this we will derive a dynamical version of the Kohn-Sham equations, which will allow us to carry
out real-time propagations of quantum systems, starting from arbitrary initial states and under the influence
of arbitrary time-dependent potentials. We will derive the formal framework of TDDFT in Section \ref{sec:IV}, and
we will discuss practical aspects and applications in Section \ref{sec:V}.

Of particular importance are situations in which the external time-dependent potential can be considered a
weak perturbation. Very often one is interested in the first-order response of the system to a perturbation,
because many spectroscopic techniques are used this regime. In particular, the linear response of a material
is directly related to its spectrum of excitations.
As we will see in Sections \ref{sec:VI} and \ref{sec:VII}, TDDFT in the linear-response regime is a very powerful approach to
calculate excitation energies and optical spectra. In fact, this is where at present the majority of TDDFT applications
are carried out at present.


\section{The formalism of TDDFT} \label{sec:IV}

\subsection{The Runge-Gross theorem} \label{subsec:IV.A}

The foundation of ground-state DFT is the Hohenberg-Kohn theorem, which we discussed in Section \ref{subsec:II.B}.
The unique 1:1 correspondence between ground-state densities and potentials makes it possible to construct
density functionals in a meaningful way, and to determine ground-state properties in principle exactly via self-consistent solution of the
Kohn-Sham equation.

For the time-dependent case we would like a similar rigorous formal foundation. But the situation is different
from the ground state, in two important ways. First, in the time-dependent case we do not have a variational minimum
principle. Secondly, the Schr\"odinger equation (\ref{III.1}) is an initial-value problem, so whatever we will prove
has to be done with a given initial state in mind.

The first to deliver an existence proof for TDDFT were Runge and Gross in 1984 \cite{Runge1984}.
They proved that if two $N$-electron systems start from the same initial state, but are subject
to two different time-dependent potentials, their respective time-dependent densities will be different.

We consider two time-dependent potentials to be {\em different} if their difference is more than just a time-dependent constant,
\begin{equation} \label{IV.1}
v(\bfr,t) - v'(\bfr,t) \ne c(t)
\end{equation}
for $t>t_0$. Otherwise they would give rise to two wave functions that differ only by a phase factor $e^{-i \alpha(t)}$, where
$d\alpha(t)/dt = c(t)$, as can easily be shown. Such purely time-dependent phase factors cancel out when one forms
expectation values of operators using Eq.  (\ref{III.5}).

The Runge-Gross theorem applies to potentials that can be expanded in a Taylor series about the initial time:
\begin{equation} \label{IV.2}
v(\bfr,t) = \sum_{k=0}^{\infty} \frac{v_k(\bfr)}{k!}  \: (t-t_0)^k \:.
\end{equation}
For such potentials, the following unique 1:1 correspondence can be proven:
\begin{equation} \label{IV.3}
v(\bfr,t)
\unitlength1cm
\begin{picture}(3.5,0.5)
\put(0.2,0.1){\vector(1,0){3}}
\put(3.2,0.1){\vector(-1,0){3}}
\put(1.0,0.25){\small unique 1:1}
\put(1.2,-0.25){\small fixed $\Psi_0$}
\end{picture}
n(\bfr,t) \:.
\end{equation}
The proof proceeds in two steps. In the first step it is established that different potentials
produce different current densities, infinitesimally later than the initial time $t_0$.
One then goes on to show that if the current densities are different, the densities must be different
as well; to prove this, the continuity equation (\ref{III.7}) is used.

Just like in ground-state DFT, the unique 1:1 correspondence (\ref{IV.3}) allows us to write the potential as a
functional of the density:
\begin{equation} \label{IV.4}
v(\bfr,t) = v[n,\Psi_0](\bfr,t) \:.
\end{equation}
Notice the formal dependence on the initial state. However, this dependence goes away if the system starts from
the ground state, i.e., $\Psi_0=\Psi\gs$: the Hohenberg-Kohn theorem then tells us that $\Psi\gs[n]$ is a functional of the
density, and $v(\bfr,t)$ can be thus written as a functional of the density only.

Since the potential can be written as a functional of the density, the time-dependent Hamiltonian becomes
a density functional as well, and hence the time-dependent wave function and all observables:
\begin{equation} \label{IV.5}
O(t) = \langle \Psi[n,\Psi_0](t) | \hat O | \Psi[n,\Psi_0](t)\rangle = O[n,\Psi_0](t) \:.
\end{equation}

\subsection{Time-dependent Kohn-Sham formalism} \label{subsec:IV.B}

\begin{figure}
\centering
\includegraphics[width=\linewidth]{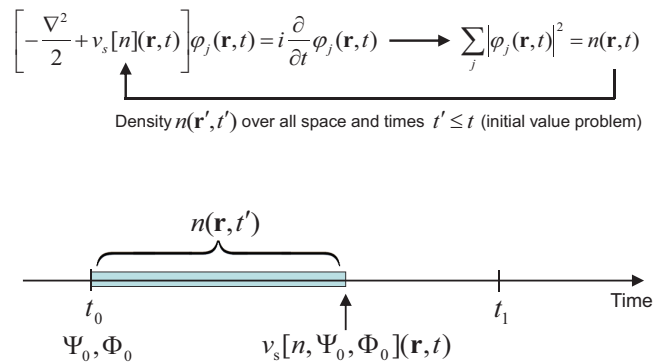}
\caption{The time-dependent Kohn-Sham equation determines the time-dependent density self-consistently between the
initial time $t_0$ and some final time $t_1$.
The xc potential at time $t$ depends on densities at times $t' \le t$, as well as on the initial states of the
interacting and of the Kohn-Sham system.} \label{fig4}
\end{figure}

The Kohn-Sham formalism (see Section \ref{subsec:II.C}) has been tremendously successful in ground-state DFT.
Its time-dependent counterpart looks very similar. The exact time-dependent density, $n(\bfr,t)$, can be
calculated from a noninteracting system with $N$ single-particle orbitals:
\begin{equation} \label{IV.6}
n(\bfr,t) = \sum_{j=1}^N |\varphi_j(\bfr,t)|^2 \:.
\end{equation}
The orbitals $\varphi_j(\bfr,t)$ satisfy the time-dependent Kohn-Sham equa\-tion:
\begin{equation} \label{IV.7}
i \frac{\partial}{\partial t} \varphi_j(\bfr,t) =
\left[ -\frac{\nabla^2}{2}  + v_s(\bfr,t)\right] \varphi_j(\bfr,t) \:,
\end{equation}
where the time-dependent effective potential is given by
\begin{equation} \label{IV.8}
v_s[n,\Psi_0,\Phi_0](\bfr,t) = v(\bfr,t) + v_{\rm H}(\bfr,t) + v_{\rm xc}[n,\Psi_0,\Phi_0](\bfr,t) \:.
\end{equation}
Here, $v(\bfr,t)$ is the time-dependent external potential, which we assume to have the form (\ref{III.4}).
The time-dependent Hartree potential,
\begin{equation}
v_{\rm H}(\bfr,t) = \int d^3r' \frac{n(\bfr',t)}{|\bfr - \bfr'|} \:,
\end{equation}
depends on the instantaneous time-dependent density only. The time-dependent xc potential formally has a
functional dependence on the density, the initial many-body state $\Psi_0$ of the exact interacting system, and
the initial state of the Kohn-Sham system $\Phi_0$. This is schematically illustrated in Fig. \ref{fig4}.

\subsection{Discussion: beyond Runge-Gross} \label{subsec:IV.C}

The Runge-Gross theorem in and by itself is sufficient to serve as the fundamental formal basis of TDDFT. However,
there are some subtle questions that it leaves unanswered and some situations that are not covered by it.
Extending the Runge-Gross theorem, or coming up with alternative proofs, has therefore been an area of
significant research activity.

This section can be skipped by readers who may be less interested in the formal details of TDDFT, and more interested
in practical aspects.

\subsubsection{$v$-representability and the van Leeuwen theorem}

An important question in ground-state DFT is the following: given a well-behaved (i.e., continuous and not singular)
mathematical function $n(\bfr)$, with $\int d^3r n(\bfr)=N$, can one always find a potential
$v_0(\bfr)$ where this $n(\bfr)$ is a ground-state density? This is known as the {\em $v$-representability} question; one distinguishes
the interacting and the noninteracting $v$-representability problem, depending on whether the given density is to be reproduced
in the physical (interacting) or in the Kohn-Sham (noninteracting) system.

Why is $v$-representability an important issue? If there exist density functions that are not $v$-representable (VR), then
the domain of the functional $E_{v_0}[n]$ would be ill defined, and one would run into formal problems in
defining functional derivatives such as in Eq. (\ref{II.17}). The $v$-representability problem in DFT is still not
fully solved, but at least we do know that all density functions on lattice systems are VR \cite{Chayes1985}
(ensemble-VR, to be precise). Fortunately, it turns out that the $v$-representability problem in ground-state DFT can
be circumvented in an elegant way with the so-called constrained search formalism \cite{Levy1979,Lieb1983}, which is,
essentially, a clever reformulation of the variational minimum principle as a search over antisymmetric $N$-particle wave functions,
so that
\begin{equation}
E_{v_0}[n]=\mbox{min}_{\Psi \to n} \langle \Psi | \hat T + \hat V_0 + \hat W | \Psi \rangle .
\end{equation}

For TDDFT the situation is different, due to a fundamental difference between the ground-state problem and the time-dependent problem:
rather than finding a ground state, TDDFT describes the time-propagation of many-body systems under the influence of
external potentials. Due to the central role the external potential plays, the $v$-representability problem
[i.e., whether there exists a $v(\bfr,t)$ for every $n(\bfr,t)$] seems unavoidable.

TDDFT is not formulated on the basis of a variational minimum principle, since there is no quantity
equivalent to the role of energy in time-dependent systems. Instead, it is possible to formulate TDDFT via
a stationary-action principle \cite{VanLeeuwen1998,VanLeeuwen2001,Vignale2008}. However, the uniqueness
of the stationary-action point remains unproven.
A rigorous time-dependent version of the constrained-search approach does not exist, despite some attempts \cite{Kohl1986,Daligault2012}.

Some progress has been made with the time-dependent $v$-representability problem for lattice systems \cite{Farzanehpour2012}.
Interestingly, in this case it can happen very easily that perfectly well-behaved lattice densities are not VR,
for well-understood reasons \cite{Li2008}.

\begin{figure}
\centering
\includegraphics[width=0.6\linewidth]{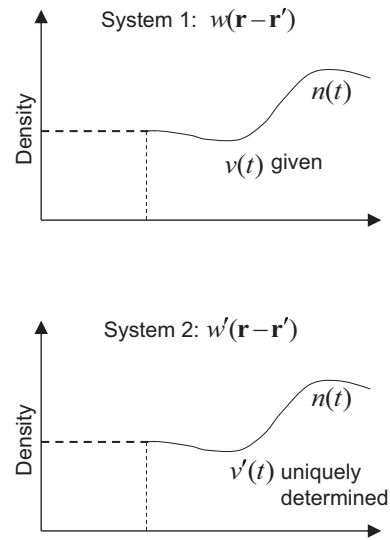}
\caption{The van Leeuwen theorem states that a time-dependent density $n(t)$ coming from a many-body system
with interaction $w(\bfr-\bfr')$ and potential  $v(\bfr,t)$ can be reproduced in a system with different interaction, $w'(\bfr - \bfr')$
and potential $v'(\bfr,t)$. The potential $v'$ is uniquely determined.} \label{fig5}
\end{figure}

The van Leeuwen theorem \cite{VanLeeuwen1999} made an important contribution towards the resolution of the $v$-representability
problem in TDDFT. It makes a statement about two many-body systems with different particle-particle interactions, $w(\bfr - \bfr')$
(system 1) and $w'(\bfr-\bfr')$ (system 2), see Fig. \ref{fig5}. If a time-dependent density $n(\bfr,t)$ is produced by an
external potential $v(\bfr,t)$ in system 1 (starting from a given initial state), then one can uniquely construct the potential
$v'(\bfr,t)$ that produces the same density in system 2 (the choice of initial state in system 2 is unique, too). There are some
restrictions on the admissible densities: they must possess a Taylor expansion in $t$  about the initial time (we denote such densities as
$t$-TE). Below, we show that this assumption can be problematic.

The van Leeuwen theorem has two important special cases. The first is that of $w=w'$, i.e., the two systems are identical.
It turns out that in this way one gains an alternative proof of the Runge-Gross theorem. The second case is $w'=0$, i.e.,
the second system is noninteracting. This establishes noninteracting $v$-representability in TDDFT, and hence provides
formal justification of the time-dependent Kohn-Sham approach.

\subsubsection{Non-Taylor-expandable densities}
The van Leeuwen theorem shows that for $t$-TE densities, one can always construct the
corresponding $t$-TE potential for the TDKS system. However, there is a subtle difference between the domain of
the van Leeuwen theorem and the Runge-Gross theorem, as the latter only requires the external
potentials to be $t$-TE [Eq. (\ref{IV.2})], but not the densities. The van Leeuwen theorem does not apply to non-$t$-TE densities,
which are allowed by the Runge-Gross theorem. Such densities are commonly considered pathological
and thus are not considered to pose any threat. However, it turns out \cite{Maitra2010,Yang2012} that the
densities of most real world systems can become non-$t$-TE, including atoms, molecules, and solids!

In the usual non-relativistic quantum mechanical description, the nuclei and electrons interact through the diverging
Coulomb potential, and the densities always have cusps at the positions of nuclei \cite{Kato1957}. The dynamics
of the system, including the time-dependent density, is determined by the time evolution operator $\hat U(t,t_0)$, which in turn
follows from the Hamiltonian $\hat H(t)$.
In the presence of space-non-analytic features such as cusps, time-non-analyticities appear
because of the kinetic energy operator $\hat T$, which is a differential operator in space.
Thus, the time-dependent density can become non-$t$-TE.

\begin{figure}
\centering
\includegraphics[width=0.8\linewidth]{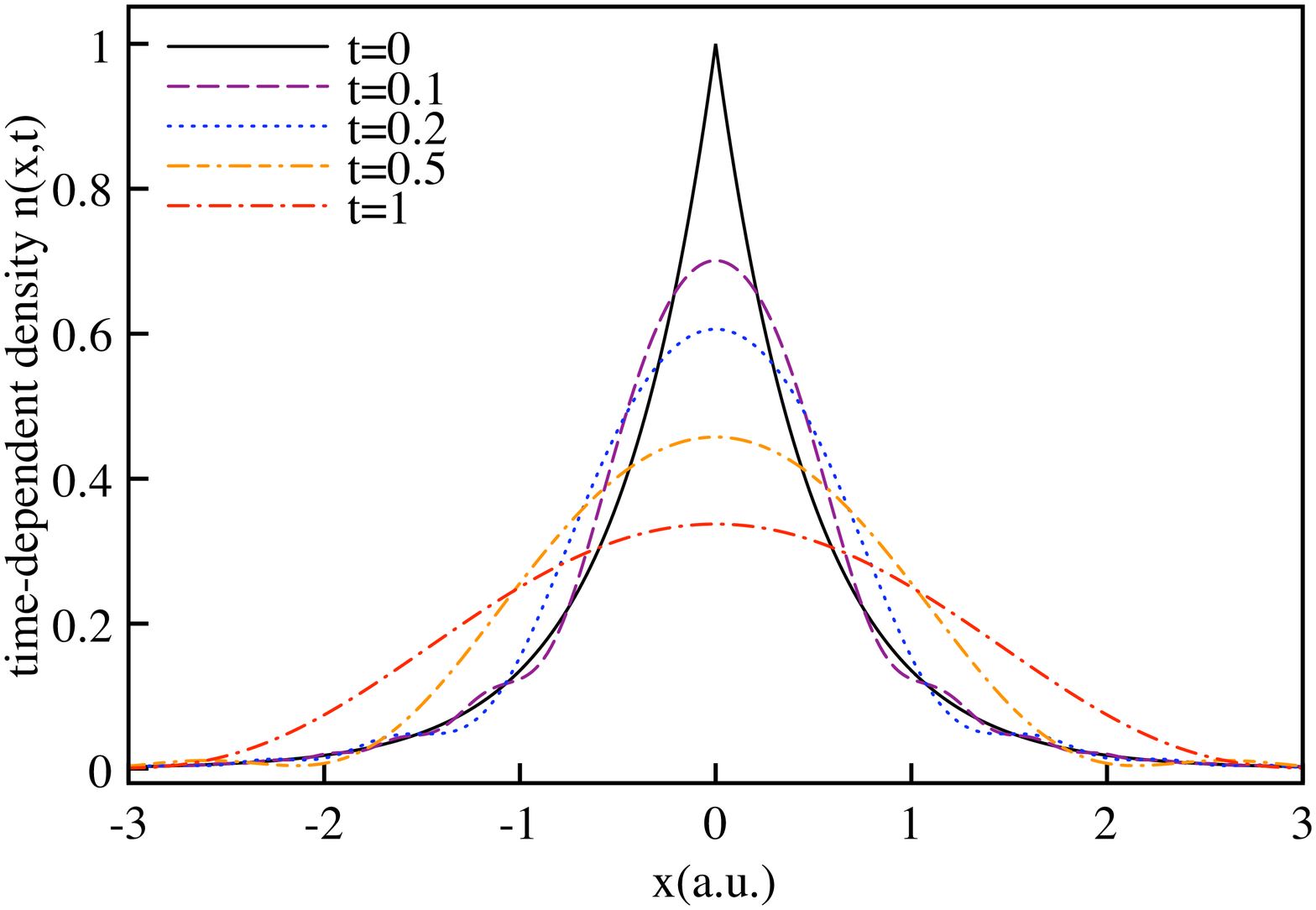}\\
\includegraphics[width=0.8\linewidth]{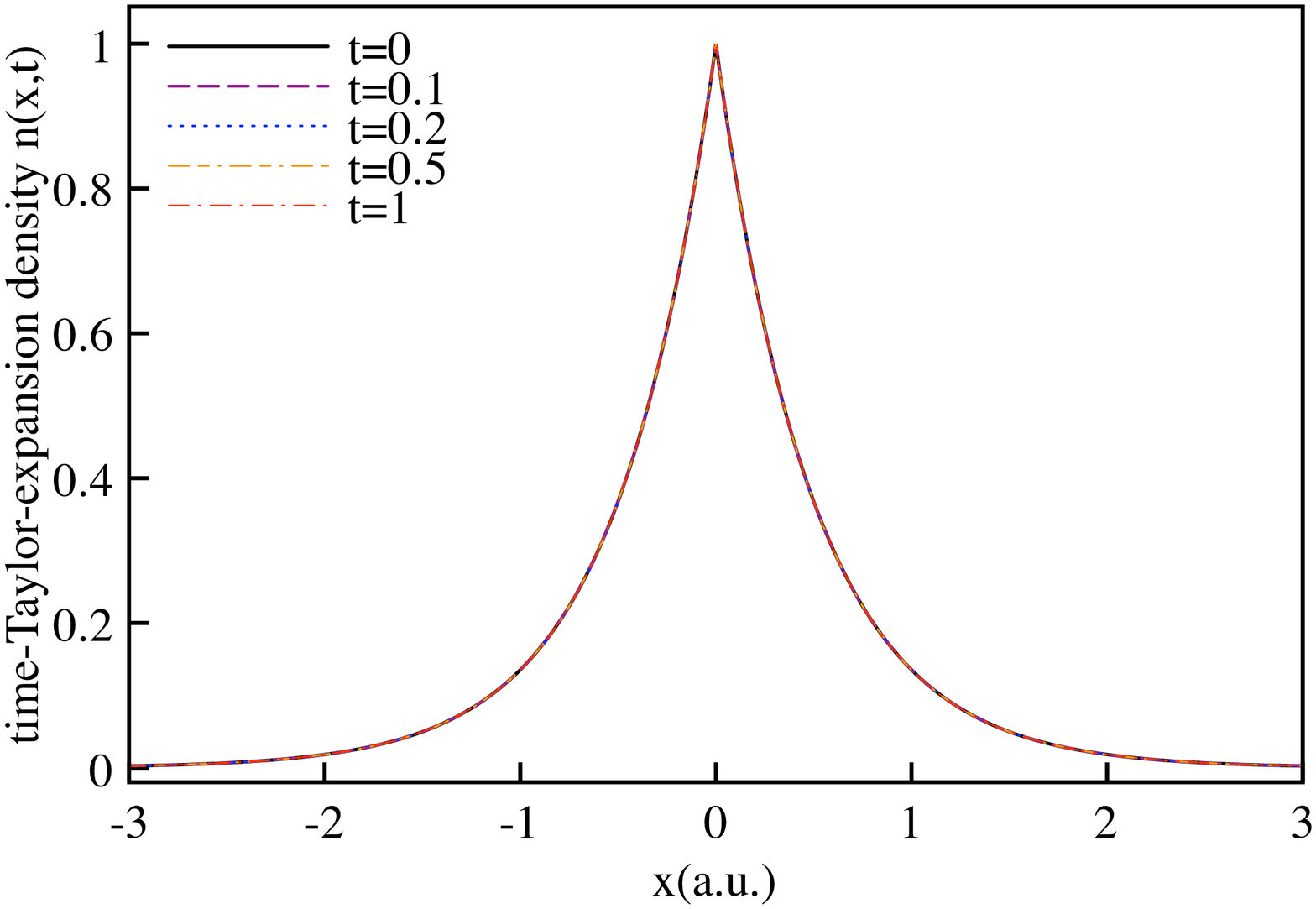}
\caption{Upper panel: time-dependent density of a 1D system with initial state
$\psi(x)=\exp(-|x|)$ propagating with no external potential. Lower panel: using a Taylor expansion in time,
the initial density remains stationary. This wrong behavior is due to the nonanalyticity of the density.}
\label{fig6}
\end{figure}

A striking example \cite{Maitra2010} demonstrating the difference between
the exact density and the $t$-TE density is shown in Fig. \ref{fig6} (the $t$-TE density is defined as the
result of applying the $t$-TE time-evolution operator on the initial state). At the initial time, a density with a cusp
is prepared, and then allowed to freely evolve in time. The upper panel of Fig. \ref{fig6} shows that the density
rapidly becomes smooth and spreads out. By contrast, if one attempts to find the time evolution by using a Taylor
expansion, the density does not move at all!

We emphasize that although the
Runge-Gross theorem is explicitly formulated for $t$-TE densities, the original proof remains valid despite the
existence of non-$t$-TE densities \cite{Yang2012}. Thus, the foundations of TDDFT remain sound.

\subsubsection{Fixed-point proofs}
Recent work on the $v$-representability problem and related questions focuses
on developing so-called fixed-point proofs \cite{Ruggenthaler2011,Ruggenthaler2012},
where the previous limitation of $t$-TE is lifted. The van Leeuwen theorem provides a way of constructing the
time-dependent external potential for a given density, if the density is $t$-TE; if applied on non-$t$-TE densities,
the constructed potential does not correspond to the exact density, but in turn reproduces the $t$-TE density \cite{Yang2012}.
The fixed-point proofs \cite{Ruggenthaler2011,Ruggenthaler2012} thus focus on explicitly showing the one-to-one
correspondence between the potential and the density. The proof starts from the equation of motion of the
density \cite{VanLeeuwen1999}:
\begin{equation}
\frac{\partial^2 n(\bfr,t)}{\partial t^2}-\nabla\cdot[n(\bfr,t)\nabla v(\bfr,t)]=q(\bfr,t).
\label{eqn:densityEOM}
\end{equation}
The density and the quantity $q$ can be seen as functionals of the potential, and thus Eq. \parref{eqn:densityEOM}
uniquely maps a potential $v_0$ to $q[v_0]$, with the density $n[v_0,\Psi_0]$ determined by $v_0$ and the initial
wave function $\Psi_0$. In another perspective, Eq. \parref{eqn:densityEOM} can also be seen as a differential
equation for the potential, when $n$ and $q$ are given. If this given density is chosen to coincide with the
initial density of the system and with its first-order time-derivative, and $q$ is chosen to be $q[v_0]$,
Eq. \parref{eqn:densityEOM} can be solved for the potential, denoted as $v_1$. Ref. \cite{Ruggenthaler2011}
proves that under mild restrictions, $v_0=v_1$, showing the mutual correspondence between the density and the
potential. The proof is strengthened by recent numerical simulations \cite{Ruggenthaler2012}.
The fixed-point proofs apply to densities confined within a finite (but arbitrarily large) space region,
and the cases of density cusps are included in a limiting sense. It is not clear as of now whether these
restrictions are general enough for the $v$-representability problem.

\subsubsection{Vector potentials and time-dependent current-DFT}

TDDFT applies to electronic many-body systems in the presence of time-dependent scalar potentials.
But there are important classes of time-dependent processes that are not included, namely, many-body systems
in time-dependent magnetic fields or under the influence of electromagnetic waves. This is obviously a very
severe omission, because this means that, strictly speaking, this precludes discussing the interaction between light
and matter! In practice, we can often get around this restriction and treat electromagnetic fields in dipole approximation, so
that TDDFT is applicable. But in the general case, to deal with vector potentials of the form $\bfA(\bfr,t)$ we need a theory that goes beyond TDDFT.

In general, a system can be under the influence of both a scalar and a vector potential, $v(\bfr,t)$ and $\bfA(\bfr,t)$.
The many-body Hamiltonian is then given by
\begin{equation}
\hat H(t) = \sum_{j=1}^N\left\{ \frac{1}{2}\left[ \frac{\nabla_j}{i} + \bfA(\bfr_j,t)\right]^2
+v(\bfr_j,t)\right\} + \hat W.
\end{equation}
The time-dependent many-body wave function associated with $\hat H(t)$ determines the density $n(\bfr,t)$ and the
current density $\bfj(\bfr,t)$. It is important to keep in mind that the current density, like any general vector field,
has a longitudinal and a transverse component,
\begin{equation}
\bfj (\bfr,t) = \bfj_L(\bfr,t) + \bfj_T(\bfr,t) \:.
\end{equation}
The longitudinal current density is related to the density via the continuity equation:
\begin{equation}
\frac{\partial }{\partial t} n(\bfr,t) = -\nabla \cdot \bfj_L(\bfr,t) \:,
\end{equation}
but the transverse component $\bfj_T(\bfr,t)$ is not determined by $n$. Hence, current densities are, in general,
not VR \cite{DAgosta2005}: if $\bfj(\bfr,t) = \bfj_L(\bfr,t) + \bfj_T(\bfr,t)$ comes from a potential $v(\bfr,t)$,
then $\bfj'(\bfr,t) = \bfj_L(\bfr,t) + \bfj'_T(\bfr,t)$ (same longitudinal but different transverse component) cannot
come from a potential $v'(\bfr,t)$, since this would violate the Runge-Gross theorem. Hence,
we need the full mapping
\begin{equation}
(v,\bfA) \leftrightarrow (n,\bfj) \:.
\end{equation}
However, this map is determined up to within a gauge transformation:
\begin{eqnarray}
v(\bfr,t) \to v(\bfr,t) - \frac{\partial}{\partial t} \Lambda(\bfr,t)
\\
\bfA(\bfr,t) \to \bfA(\bfr,t) + \nabla \Lambda(\bfr,t) \:,
\end{eqnarray}
where $\Lambda(\bfr,t)$ is an arbitrary (but well-behaved) gauge function which vanishes at the initial time.
Often, one chooses the gauge function in such a way that the scalar potential vanishes.

Ghosh and Dhara \cite{Ghosh1988} were the first to give a formal proof of time-dependent current-DFT (TDCDFT). More recently,
an alternative existence proof of TDCDFT, in the spirit of the van Leeuwen theorem, was provided by Vignale \cite{Vignale2004}.
TDCDFT on lattice systems was discussed by Tokatly \cite{Tokatly2011}.
The time-dependent Kohn-Sham equation in TDCDFT becomes
\begin{equation}
i \frac{\partial }{\partial t} \varphi_j(\bfr,t)
= \left\{\frac{1}{2} \left[\frac{\nabla}{i} + \bfA_s(\bfr,t)\right]^2 + v_s(\bfr,t)\right\} \varphi_j(\bfr,t)\:,
\end{equation}
where the effective scalar potential, as before, is given by Eq. (\ref{IV.8}), and the effective vector potential is
\begin{equation}
\bfA_s(\bfr,t) = \bfA(\bfr,t) + \bfA_{\rm xc}(\bfr,t) \:.
\end{equation}
Notice that the effective vector potential does not contain a Hartree-like term due to induced currents, since this would
be relativistically small. The gauge-invariant physical current density is given by
\begin{equation}
\bfj(\bfr,t) = n(\bfr,t) \bfA_s(\bfr,t) + \frac{1}{i} \sum_{j=1}^N
\Im \left[\varphi_j^*(\bfr,t) \nabla \varphi_j(\bfr,t)\right] .
\end{equation}

Let us summarize the key points of TDCDFT:

\begin{enumerate}

\item TDCDFT overcomes formal limitations of TDDFT, allowing treatment of electromagnetic waves and general vector potentials
and time-varying magnetic fields. However, electromagnetic waves are usually treated in dipole approximation, so one
rarely makes use of TDCDFT in this way.

\item The Runge-Gross theorem of TDDFT has been proved for finite systems, where the density vanishes at infinity.
However, it also works for periodic systems \cite{Maitra2003}, provided the external potential is also periodic.
The Runge-Gross theorem does {\em not} apply when a uniform homogeneous field acts on a periodic system.
This case, however, is formally included in TDCDFT \cite{Vignale2004}.

\item TDCDFT can be very useful in situations that could, in principle, be fully described with TDDFT;
using the current as basic variable, rather than the density, can make it easier to develop approximations
for dynamical xc effects \cite{Vignale1996,Vignale1997}.

\end{enumerate}

\section{Practical aspects} \label{sec:V}

To apply TDDFT in practice requires the following
considerations:

\begin{itemize}

\item a suitable approximation for the time-dependent xc potential needs to be found;

\item the time-dependent Kohn-Sham equations need to be solved numerically;

\item the physical observables of interest need to be obtained from the time-dependent density.

\end{itemize}

Each of these points has  its own challenges.
We shall now address them individually, including some examples.

\subsection{The time-dependent xc potential} \label{subsec:V.A}

As we said in Section \ref{subsec:IV.B}, the time-dependent xc potential is formally a functional
of the time-dependent density as well as the initial states, $v_{\rm xc}[n,\Psi_0,\Phi_0](\bfr,t)$.
In practice, one is usually interested in situations where the system is initially in the ground state.
If this is the case, things simplify considerably: thanks to the Hohenberg-Kohn theorem of ground-state DFT,
the initial states become functionals of the initial (ground-state) density, and the xc functional
can be written as a density functional only, $v_{\rm xc}[n](\bfr,t)$.

However, the density-dependence of the xc potential is complicated and nonlocal:
the xc potential at space-time point $(\bfr,t)$ depends on densities at all other points in space
and at all previous times, $n(\bfr',t')$, where $t'\le t$ (the potential cannot depend
on densities in the future---this would violate the fundamental principle of causality).

The most widely used approximation for the xc potential is the {\em adiabatic approximation}:
\begin{equation}\label{V.1a}
v_{\rm xc}^{\rm A}(\bfr,t)=v_{\rm xc}^{\rm gs}[n_0](\bfr)|_{n_0(\bfr)=n(\bfr,t)},
\end{equation}
where $v_{\rm xc}^{\rm gs}$, the ground-state xc potential defined in Eq. \parref{II.17},
is evaluated at the instantaneous time-dependent density.
Eq. \parref{V.1a} becomes exact for an infinitely slowly varying system
which is in its ground state for any time. In practice, this is of course not the case
(unless one considers a time-dependent system which just sits there in its ground state, doing nothing).

One of the most important questions in TDDFT is under what circumstances the adiabatic approximation
works well. Numerical studies \cite{Ullrich2006a,Thiele2008,Thiele2009} demonstrate that
the adiabatic approximation may break down if the system undergoes very rapid changes, but
it turns out that the adiabatic approximation still works surprisingly well in many cases.
This will be further addressed below when we discuss the calculation of excitation energies.

As of today, very few applications of TDDFT have been carried out with nonadiabatic,
explicitly memory-dependent xc functionals \cite{Wijewardane2005,Wijewardane2008,Kurzweil2006,Kurzweil2008}.
Due to its simplicity, the overwhelming majority of time-dependent Kohn-Sham calculations
use the adiabatic LDA (ALDA),
\begin{equation}
v_{\rm xc}^{\rm ALDA}(\bfr,t) = v_{\rm xc}^{\rm LDA}(n(\bfr,t)) \:,
\end{equation}
or any adiabatic GGA defined in a similar way, by replacing the ground-state density with the
instantaneous time-dependent density.

\subsection{Observables} \label{subsec:V.B}

In Section \ref{subsec:IV.A} we showed that all physical observables are formally functionals of the time-dependent
density, see Eq. (\ref{IV.5}). TDDFT gives, in principle, the exact time-dependent density $n(\bfr,t)$, and all
quantities of interest must be obtained from it. Some observables are easily calculated in this way,
but other are not. We will now give examples of both kinds.

\subsubsection{Easy observables}

The easiest observable is the density itself, which shows how electrons move during any time-dependent process.
This is certainly useful for visualizing molecular geometries
or structural changes during chemical reactions or photoinduced processes, but does not reveal important quantum mechanical features
such as atomic shell structure, covalent molecular bonds, or lone pairs. Such information can be gained from
a convenient visualization tool known as
the time-dependent electron localization function (TDELF) \cite{Burnus2005}.
The TDELF is defined as a positive quantity with a magnitude between zero and one:
\begin{equation} \label{5.TDELF}
f_{\rm ELF}(\bfr,t) = \frac{1}{1 + [D_\sigma(\bfr,t)/D_\sigma^0(\bfr,t)]^2} \:.
\end{equation}
The quantity
\begin{equation}
D_\sigma(\bfr,t) = \tau_\sigma(\bfr,t) - \frac{|\nabla n_\sigma(\bfr,t)|^2}{8n_\sigma(\bfr,t)}
- \frac{|\bfj_\sigma(\bfr,t)|^2}{2n_\sigma(\bfr,t)}
\end{equation}
is a measure of the probability of finding an electron in the vicinity
of another electron of the same spin $\sigma$ at $(\bfr,t)$.
Clearly, $D_\sigma(\bfr,t)$  is not an explicit density functional,
but it is expressed in terms of the density, the current,
and the orbitals via the kinetic-energy density $\tau_\sigma(\bfr,t)
= \frac{1}{2}\sum_{j=1}^{N_\sigma} |\nabla \varphi_{j\sigma}(\bfr,t)|^2$.
$D_\sigma^0$ in Eq. (\ref{5.TDELF})
is given by the kinetic-energy density of the homogeneous electron liquid:
\begin{equation}
D_\sigma^0(\bfr,t) = \frac{3}{10}(6\pi^2)^{3/2} n_\sigma^{5/3}(\bfr,t) = \tau_\sigma^{h}(\bfr,t) \:.
\end{equation}

The time propagation is unitary, so the total norm is conserved; but to describe ionization or charge transfer
processes, it is often of interest to obtain the number of electrons that escape from a given
spatial region $\cal V$:
\begin{equation} \label{4.2.1}
N_{\rm esc}(t) = N - \int_{\cal V} d^3\: n(\bfr,t) \:.
\end{equation}
Here, $\cal V$ can be thought of as a ``box'' that surrounds the entire system (in case we wish to calculate
ionization rates of atoms or molecules), or it could be a part of a larger molecule or part of a unit cell of a
periodic solid.

Another easy class of observables are moments of the density, such as the dipole moment:
\begin{equation} \label{4.2.2}
{\bf d}(t) = \int d^3r \: \bfr n(\bfr,t) \:.
\end{equation}
The dipole moment can be considered directly, i.e., in real time, to study the behavior of charge-density oscillations.
Alternatively, it can be Fourier transformed to yield the dipole power spectrum $|d(\omega)|^2$ or related
observable quantities such as the photoabsorption cross section.

Higher moments of the density, such as the quadrupole moment, can be calculated just as easily, but are
less frequently considered.

\subsubsection{Difficult observables}

Equation (\ref{4.2.1}) gives the total number of escaped electrons, which in general can be nonintegral.
For instance, if we consider an atom in a laser field, a value of $N_{\rm esc}=0.5$ would indicate that
on average half an electron has been removed. In reality there are of course no ``half-electrons'', so we
have to interpret this result in a probabilistic sense: it could for instance mean that there is 50\% probability that
the atom is singly ionized, and 50\% probability that it is not ionized; other scenarios, involving
double ionization, are also possible. The probabilities to find an atom or molecule in a certain charge state $+m$
can be defined as follows \cite{Ullrich2000}:
\begin{eqnarray}
P^0(t) &=& \int\limits_{\cal V} d^3r_1 \ldots \int\limits_{\cal V} d^3r_N \:|\Psi(\bfr_1,\ldots,\bfr_N,t)|^2
\\
P^{+1}(t) &=& \int\limits_{\overline{\cal V}}\!  d^3r_1 \! \int\limits_{\cal V} \! d^3r_2 \ldots \! \int\limits_{\cal V} d^3r_N
|\Psi(\bfr_1,\ldots,\bfr_N,t)|^2 \hspace{6mm}
\end{eqnarray}
and similar for all other $P^{+m}(t)$. Here $\overline{\cal V}$ denotes all space outside the integration box $\cal V$
surrounding the system. The ion probabilities are defined in terms of the full many-body wave function $\Psi(t)$,
which is a density functional according to the Runge-Gross theorem; but it is not possible to extract the
ion probabilities $P^{+m}(t)$ directly from the density in an elementary way.

Since the full wave function is
prohibitively expensive to deal with, a pragmatic solution is to replace $\Psi(t)$ by the Kohn-Sham Slater determinant $\Phi(t)$,
in spite of the fact that the latter has no rigorous physical meaning. One then obtains the Kohn-Sham
ion probabilities
\begin{eqnarray}
P_s^0(t) &=& N_1(t) N_2(t) \ldots N_N(t)
\\
P_s^{+1}(t) &=& \sum_{j=1}^N N_1(t)\ldots N_{j-1}(t)\big(1-N_j(t)\big)
\nonumber\\
&& \times N_{j+1}(t) \ldots N_N(t)
\end{eqnarray}
and similar for all other $P_s^{+m}(t)$, where
\begin{equation}
N_j(t) = \int\limits_{\cal V} d^3r |\varphi_j(\bfr,t)|^2 \:.
\end{equation}
The Kohn-Sham ion probabilities are easily obtained from the orbitals; but apart from certain
limiting cases  \cite{Ullrich2000}, they are have no rigorous physical meaning \cite{Lappas1998,Wilken2006}.
Here are some other examples of difficult observables:

{\em Photoelectron spectra}. The photoelectron kinetic energy distribution spectrum is formally defined as
\begin{equation}
P(E)dE=\lim_{t\to\infty}\sum_{k=1}^N|\langle \Psi_E^k|\Psi(t)\rangle|^2dE \:,
\end{equation}
where $|\Psi_E^k\rangle$ is a many-body eigenstate with $k$ electron in the continuum and total kinetic energy $E$ of the
continuum electrons. There are approximate ways of calculating photoelectron spectra from the density or from the Kohn-Sham orbitals
\cite{Pohl2000,Veniard2003,DeGiovannini2012}.

{\em State-to-state transition probabilities.} The S-matrix describes the transition between two states:
\begin{equation}
S_{i,f} = \lim_{t\to \infty} \langle \Psi_f | \Psi(t)\rangle \:,
\end{equation}
for given initial and final many-body states $\Psi_i$ and $\Psi_f$. To get the S-matrix
from the density, a cumbersome implicit read-out procedure was proposed \cite{Rohringer2006}.

{\em Momentum distributions}. Ion recoil momenta are of great interest in high-intense field or
scattering experiments. The problem is formally similar to the problem of calculating ion probabilities
from the density, and in principle requires the full wave function in momentum space. The Kohn-Sham momentum
distributions can be taken as approximation, without formal justification \cite{Wilken2007}.

{\em Transition density matrix}. The transition density matrix is a quantity that is defined in the
linear response regime. As the name indicates, it refers to a specific excitation of the system (typically,
a large molecular system), and maps the distribution and coherences of the excited electron and the associated hole.
In particular, the transition density matrix is useful to visualize excitonic effects. There is no
easy way to obtain it directly from the density; the best we can do is to construct the transition density
matrix from Kohn-Sham orbitals \cite{Li2011}.

All the above examples have in common that they are explicit expressions of the many-body wave function, or of the
$N$-body density matrix, but can only be implicitly expressed as density functionals.
One can get approximate results by replacing the full many-body wave function with the Kohn-Sham
Slater determinant $\Phi(t)$, but there is no guarantee that this will give good results.

\subsection{Applications}  \label{subsec:V.C}

Real-time TDDFT  has been implemented in several
computer codes, most notably the open-source code {\tt octopus} \cite{Castro2006,Andrade2012}.
A TDDFT code must deal with two basic numerical tasks:
(i) The Kohn-Sham orbitals of the system, and its density, must be represented in space.
This can be done either with a suitable basis, or on a spatial grid using finite-element
or finite-difference discretization schemes ({\tt octopus}
uses the latter).
(ii) Time must be discretized as well, and the time-dependent Kohn-Sham equations
are propagated forward in time, step by step, ensuring norm conservation.

Let us say a few words about the time propagation.
Suppose we know the Kohn-Sham orbitals up until some time $\tau_n$. The orbitals at the next time
step, $\tau_{n+1} = \tau_n + \Delta \tau$, can then formally be written as
\begin{equation} \label{timeprop}
\varphi_j(\tau_n + \Delta \tau) = \hat U(\tau_n + \Delta \tau,\tau_n)\varphi_j(\tau_n)\:,
\end{equation}
where $\hat U(\tau_n + \Delta \tau,\tau_n)$ is the time evolution operator which propagates
the orbitals one time step $\Delta \tau$ forward.
If $\Delta \tau$ is sufficiently small, we can approximate $\hat U$ by
\begin{equation}
\hat U (\tau_n + \Delta \tau,\tau_n) \approx e^{-i \hat H_s(\tau_n + \Delta \tau/2) \Delta \tau} \:,
\end{equation}
where $H_s(\tau_n + \Delta \tau/2)$ is the time-dependent Kohn-Sham Hamiltonian evaluated midway between
the two time steps (in practice, this requires a so-called predictor-corrector scheme \cite{Ullrich2012}).
The time propagation (\ref{timeprop}) can be numerically implemented in various ways \cite{Castro2004};
an example is the Crank-Nicholson algorithm:
\begin{equation}
e^{-i\hat H_s \Delta \tau} \approx \frac{1-i\hat H_s \Delta \tau/2}{1+i\hat H_s \Delta \tau/2} \:,
\end{equation}
which is correct to order $(\Delta \tau)^2$ and unitary (hence, the norm of the wave functions is conserved).
This converts the time-dependent Kohn-Sham equations into a set of linear equations that can
be numerically solved.

The applications of real-time TDDFT can be roughly divided into two categories, related to the two scenarios
we discussed in Section \ref{sec:III}.

In the first class of applications, the system is initially prepared
in a nonequilibrium state through a sudden switching or a short impulsive excitation, and then allowed
to propagate freely in time \cite{Yabana1996,Yabana1999,Yabana2006}. The initial perturbation is kept weak in order to avoid any nonlinear effects,
but it is spectrally broad and hence triggers a dynamical behavior of the system in which essentially the
entire range of excitations participates. The time-dependent dipole moment ${\bf d}(t)$, Eq. (\ref{4.2.2}),
is calculated over a certain time span, and Fourier transformation yields the optical spectrum of the system.
The time-propagation method has certain advantages especially for large systems \cite{Marques2003,Takimoto2007,Botti2009,Spallanzani2009}
and metallic clusters \cite{Calvayrac2000},
but is less frequently used for low-lying excitations of smaller molecules.
Below, in Section \ref{sec:VI}, we will discuss an alternative way of calculating excitation energies.

\begin{figure}
\centering
\includegraphics[width=\linewidth]{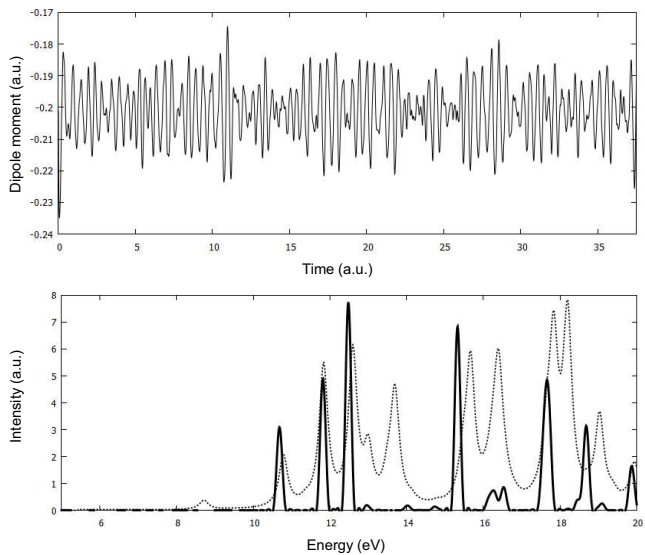}
\caption{Time-dependent Kohn-Sham calculation for a $\rm CO_2$ molecule. Top: time-dependent dipole moment $d(t)$
induced by an initial ``kick''. Bottom: dipole spectrum, obtained by Fourier
transforming $d(t)$ (full line), compared with the spectrum obtained from linear-response TDDFT (thin line).} \label{figCO2}
\end{figure}

Figure \ref{figCO2} shows an example of such a calculation for the $\rm CO_2$ molecule. The optical absorption spectrum,
obtained by Fourier transforming the time-dependent dipole moment, agrees well with a spectrum that is obtained
from linear-response TDDFT (we will discuss this approach in the following Section). Both spectra, in turn,
agree well with experiment \cite{Hubin1988}.

The second class of applications is in the nonlinear regime, and deals with
systems that are subject to strong excitations such as high-intensity laser pulses or
collisions with fast, highly charged ionic projectiles. The response following such excitations
can be highly nonlinear and far beyond any treatment using perturbative methods. Propagation of
the time-dependent Kohn-Sham equations yields the response to all orders, in principle exactly,
including collective many-body effects. Quantities of interest include easy observables
such as total ionization yields and high-harmonic generation spectra, and difficult observables
such as photoelectron spectra, ion probabilities, or momentum distributions.

\begin{figure}
\centering
\includegraphics[width=\linewidth]{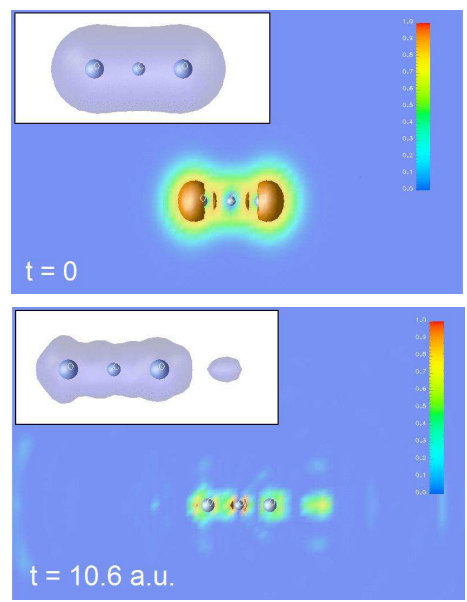}
\caption{Two snapshots of the time-dependent electron localization function (TDELF) for a $\rm CO_2$ molecule,
excited by a laser pulse of photon energy 20 eV and intensity $1.2\times 10^{15} \: \rm W/cm^2$. Insets: density isosurfaces. } \label{figCO2_strong}
\end{figure}

Figure \ref{figCO2_strong} shows an example. A $\rm CO_2$ molecule
is hit with a very short, high-intensity laser pulse which deposits a large amount of excitation energy in a very short time.
The snapshot at $t=10.6$ a.u. (1 a.u. equals 24 attoseconds) shows how a packet of density flies off, and the remaining density is strongly distorted.
The TDELF, Eq. (\ref{5.TDELF}), illustrates how the electronic orbitals have become extremely diffuse, and the bonds
are essentially destroyed, which will cause the molecule to break up.

TDDFT calculations for strong excitations have been carried out over the past two decades for a variety of atomic and
molecular systems \cite{Ullrich1997,Ullrich1997b,Lappas1998,Tong1998,Chu2001b,Nguyen2004,Wilken2006,Wilken2007,PenkaFowe2010}
(see \cite{Ullrich2012b} for a review). An intriguing question is whether it is possible to design the excitation
(i.e., the laser intensity, pulse shape, and spectral composition) in such as way that a specific control goal can be
achieved. The formal framework of TDDFT and optimal control has been worked out \cite{Werschnik2007,Castro2012},
but some of the more interesting control goals may be difficult to achieve with standard (adiabatic) TDDFT approaches
\cite{Ruggenthaler2009,Fuks2011,Raghunathan2011,Raghunathan2012a,Raghunathan2012b}.

\section{TDDFT and linear response} \label{sec:VI}

\subsection{Formalism} \label{subsec:VI.A}

In many situations of practical interest, systems are subjected to small perturbations and hence do not
deviate strongly from their initial state. This happens in most applications of spectroscopy, where the response to a
weak probe is used to determine the spectral properties of a system. In this case, it is not necessary
to seek a fully-fledged solution of the time-dependent Schr\"odinger or Kohn-Sham equations (although
this would yield the desired information, too, as we have seen in Fig. \ref{figCO2}). Instead, one can use perturbation theory.
The goal of {\em linear-response theory} is to directly calculate the change of a certain variable or observable
to first order in the perturbation, without calculating the change of the wave function.  For us, the most
important example is the linear density response.

We consider the case where the system is initially in the ground state and
a time-dependent potential is switched on at time $t_0$, see Eq. (\ref{III.4}). Now, however,
$v_1(\bfr,t)$ is treated as a small perturbation. This perturbation will cause some (small) time-dependent
changes in the system, and the density will become time-dependent. We expand it as follows:
\begin{equation} \label{V.1}
n(\bfr,t) = n_0(\bfr) + n_1(\bfr,t) + n_2(\bfr,t) + \ldots .
\end{equation}
Here, $n_0$ is the ground-state density, $n_1$ is the linear density response (the first-order change in density
induced by the perturbation $v_1$), $n_2$ is the second-order density response (quadratic in the perturbation $v_1$),
and there will be higher-order terms which we have not explicitly indicated. If the perturbation is small, the linear
density response dominates over all higher-order terms in the expansion (\ref{V.1}). On the other hand, if the perturbation
is strong, a perturbation expansion may not even converge! In that case it makes more sense to solve the Schr\"odinger (or Kohn-Sham)
equations instead. Notice that all contributions
to the density response integrate to zero, e.g.,
$\int d^3r \, n_1(\bfr,t) = 0$, due to norm conservation.

The linear density response can be formally written as
\begin{equation} \label{V.2}
n_1(\bfr,t) = \int_{-\infty}^\infty dt' \int d^3r' \chi(\bfr,t,\bfr',t') v_1(\bfr',t') \:.
\end{equation}
Here, $\chi(\bfr,\bfr',t-t')$ is the density-density response function, defined as \cite{Ullrich2012,GiulianiVignale}
\begin{equation} \label{V.3}
\chi(\bfr,t,\bfr',t') = -i\theta(t-t') \langle \Psi\gs|[\hat n(\bfr,t-t'),\hat n(\bfr')]|\Psi\gs\rangle \:.
\end{equation}
The step function $\theta(t-t')$ ensures that the response is {\em causal}, i.e.,  the response comes after the perturbation.
Equation (\ref{V.3}) shows that the response function is obtained from the many-body ground state $\Psi\gs$,
involving a commutator of density operators (in interaction representation). Hence, via the Hohenberg-Kohn theorem,
it is formally a functional of the ground-state density, $\chi[n_0]$. Usually, one is more interested
in the frequency-dependent response than in the real-time response:
\begin{equation} \label{V.4}
n_1(\bfr,\omega) =  \int d^3r' \chi(\bfr,\bfr',\omega) v_1(\bfr',\omega) \:.
\end{equation}
The Fourier transform of the response function (\ref{V.3}) can be written in the following form, known as
the Lehmann representation \cite{Ullrich2012,GiulianiVignale}:
\begin{eqnarray} \label{V.D.5}
\chi(\bfr,\bfr',\omega) &=& \sum_{n=1}^\infty \bigg\{
\frac{\langle \Psi\gs | \hat n(\bfr) | \Psi_n\rangle \langle \Psi_n | \hat n(\bfr') | \Psi\gs\rangle}
{\omega - \Omega_n + i\eta}
\nonumber\\
&& \hspace{5mm}-
\frac{\langle \Psi\gs | \hat n(\bfr') | \Psi_n\rangle \langle \Psi_n | \hat n(\bfr) | \Psi\gs\rangle}
{\omega + \Omega_n + i\eta} \bigg\}, \hspace{3mm}
\end{eqnarray}
where the limit $\eta \to 0^+$ is understood. Here,
\begin{equation} \label{V.6}
\Omega_n = E_n - E_0
\end{equation}
is the $n$th excitation energy of the many-body system. This shows explicitly that the response function
has poles at the exact excitation energies of the system. This makes sense: if we apply a perturbation $v_1(\bfr,\omega)$
whose frequency matches one of the excitation energies, the response of the system is very large (we see a peak in the spectrum).

If we knew the response function $\chi$ of the many-body system, calculating the density response would be
easy and straightforward: all we have to do is evaluate expression (\ref{V.4}). From the density response,
spectroscopic observables of interest can then be calculated. For instance, one often considers a monochromatic dipole field along,
say, the $z$ direction,
\begin{equation} \label{V.7}
v_1(\bfr,t) = {\cal E} z \sin(\omega t)\:.
\end{equation}
The dynamic dipole polarizability follows as
\begin{equation} \label{V.8}
\alpha(\omega) = -\frac{2}{\cal E} \int d^3r \: z n_1(\bfr,\omega)\:,
\end{equation}
and the photoabsorption cross section $\sigma(\omega)$ is given by
\begin{equation} \label{V.9}
\sigma(\omega) = \frac{4\pi\omega}{c} \: \Im \alpha(\omega) \:.
\end{equation}
In TDDFT, the linear density response can be calculated, in principle exactly, as the response
of the noninteracting Kohn-Sham system to an {\em effective} perturbation \cite{Gross1985}:
\begin{equation} \label{V.10}
n_1(\bfr,t) = \int dt'\int d^3r' \chi_s(\bfr,t,\bfr',t') v_{1s}(\bfr',t') \:.
\end{equation}
Here, $\chi_s(\bfr,\bfr',t-t')$ is the density-density response function of the Kohn-Sham system.
The effective perturbation is given as the sum of the real external perturbation plus the linearized Hartree
and xc potentials:
\begin{eqnarray} \label{V.11}
v_{s1}(\bfr,t) &=& v_1(\bfr,t) + \int d^3r'\:  \frac{n_1(\bfr',t)}{|\bfr - \bfr'|}
\nonumber\\
&+& \int dt' \int d^3r' f_{\rm xc}(\bfr,t,\bfr',t') n_1(\bfr',t') \:.
\end{eqnarray}
The so-called {\em xc kernel} is defined as the functional derivative of the time-dependent xc potential
with respect to the time-dependent density, evaluated at the ground-state density:
\begin{equation} \label{V.12}
f_{\rm xc}(\bfr,t,\bfr',t') = \left. \frac{\delta v_{\rm xc}[n](\bfr,t)}{\delta n(\bfr',t')}\right|_{n_0(\bfr)} \:.
\end{equation}
The effective perturbation (\ref{V.11}) depends on the density response, so the TDDFT response equation
(\ref{V.10}) has to be solved self-consistently. Again, we are usually more interested in the frequency-dependent
response, given by
\begin{equation} \label{V.13}
n_1(\bfr,\omega) = \int d^3r' \chi_s(\bfr,\bfr',\omega) v_{1s}(\bfr',\omega) \:,
\end{equation}
and
\begin{eqnarray} \label{V.14}
v_{s1}(\bfr,\omega) &=& v_1(\bfr,\omega)
\\
&+& \int d^3r'\left\{ \frac{1}{|\bfr - \bfr'|}
+ f_{\rm xc}(\bfr,\bfr',\omega)\right\} n_1(\bfr',\omega) \:. \nonumber
\end{eqnarray}
The frequency-dependent xc kernel is the Fourier transform of $f_{\rm xc}(\bfr,t,\bfr',t')$ with respect to $(t-t')$.

The Kohn-Sham response function is given by
\begin{equation}\label{V.15}
\chi_s(\bfr,\bfr',\omega) = \sum_{j,k=1}^\infty (f_k - f_j)
\frac{\varphi_j(\bfr) \varphi_k^*(\bfr) \varphi_j^*(\bfr') \varphi_k(\bfr')}{\omega - \omega_{jk} + i\eta} \:,
\end{equation}
where $f_j$ and $f_k$ are occupation numbers referring to the configuration of the Kohn-Sham ground state
(1 for occupied and 0 for empty Kohn-Sham orbitals), and the $\omega_{jk}$ are defined  as
\begin{equation} \label{V.16}
\omega_{jk} = \varepsilon_j - \varepsilon_k \:.
\end{equation}
Thus, $\chi_s(\bfr,\bfr',\omega)$ has poles at the excitation energies of the noninteracting Kohn-Sham system.
Naively, one might conclude from this that the TDDFT linear response must be wrong, since it contains a response
function with the wrong pole structure (we pointed out above that the exact response function has poles at the
exact excitation energies $\Omega_n$). The resolution to this apparent contradiction lies in the self-consistent
nature of the TDDFT response equation, which ``cancels out'' the wrong poles and restores the correct poles of the many-body system.

The TDDFT linear-response formalism can be generalized to a spin-dependent form. The response equation is then given by
\begin{equation} \label{V.10a}
n_{1\sigma}(\bfr,t) = \sum_{\sigma'}\int dt'\int d^3r' \chi_{s\sigma\sigma'}(\bfr,t,\bfr',t') v_{1s\sigma}(\bfr',t') \:,
\end{equation}
where the Kohn-Sham response function is diagonal in the spin index:
\begin{eqnarray}\label{V.15a}
\chi_{s\sigma\sigma'}(\bfr,\bfr',\omega) &=& \delta_{\sigma \sigma'}\sum_{j,k=1}^\infty (f_{k\sigma} - f_{j\sigma})
\nonumber\\
&\times&
\frac{\varphi_{j\sigma}(\bfr) \varphi_{k\sigma}^*(\bfr) \varphi_{j\sigma}^*(\bfr') \varphi_{k\sigma}(\bfr')}
{\omega - \omega_{jk\sigma} + i\eta} \:,
\end{eqnarray}
and $\omega_{jk\sigma} = \varepsilon_{k\sigma} - \varepsilon_{j\sigma}$. The effective perturbation is
\begin{eqnarray} \label{V.14a}
v_{s1\sigma}(\bfr,\omega) &=& v_{1\sigma}(\bfr,\omega)
+ \sum_{\sigma'} \int d^3r' \bigg\{ \frac{1}{|\bfr - \bfr'|}
\nonumber\\
&&
+ f_{\rm xc\sigma\sigma'}(\bfr,\bfr',\omega)\bigg\} n_{1\sigma'}(\bfr',\omega) \:,
\end{eqnarray}
featuring the spin-dependent xc kernel $f_{\rm xc \sigma\sigma'}$.


\subsection{How to calculate excitation energies} \label{subsec:VI.B}

The excitation energies of a many-body system are defined as the differences between the ground-state energy $E_0$ and
the energies of higher-lying eigenstates, $E_n$, see Eq. (\ref{V.6}). In other words, they are obtained by comparing the
energies of stationary states. Why, then, would one want to use a time-dependent approach such as TDDFT? Isn't that
unnecessarily complicated?

It helps to think of an excitation in a different way, namely, as a dynamical process where the system transitions between
two eigenstates; the excitation energy then corresponds to a characteristic frequency, which describes the
rearrangements of probability density during the transition process. In other words, each excitation corresponds to
a characteristic {\em eigenmode} of the interacting $N$-electron system.

The concept of electronic eigenmodes has a familiar analog in classical mechanics \cite{Landau}. A system of $s$ coupled
oscillators carrying out small oscillations is described by the homogeneous linear system of equations
\begin{equation} \label{V.17}
\sum_{j=1}^s(k_{ij}-\Omega^2 m_{ij})A_j = 0 \:, \quad i=1,\ldots,s,
\end{equation}
where the matrices $k_{ij}$ and $m_{ij}$ determine the potential and kinetic energy of the system, respectively:
\begin{eqnarray} \label{V.18}
U&=& \frac{1}{2} \sum_{ij}^s k_{ij} q_i q_j\\
T &=& \frac{1}{2}\sum_{ij}^s m_{ij} q_i q_j \label{V.19}
\end{eqnarray}
(the $q_j$ are generalized coordinates). Clearly, $k_{ij}$ and $m_{ij}$ generalize the concept of
spring constant and mass of a simple harmonic oscillator. The solutions of Eq. (\ref{V.17})
are obtained by finding the roots of the determinant,
\begin{equation} \label{V.20}
\mbox{det}|k_{ij} - \Omega^2 m_{ij}|=0 \:.
\end{equation}
The $s$ solutions $\Omega_\alpha^2$, $\alpha=1,\ldots,s$, are the eigenfrequencies of the system, and
the associated eigenvectors $A_{j \alpha}$ indicate the profile of the eigenmode, and can be used
to determine the normal modes of the system.

It turns out that calculating excitation energies with TDDFT is very similar to describing
the small oscillations of a classical system. Starting point is the TDDFT response equation, Eq. (\ref{V.13}),
but without any external perturbation:
\begin{equation} \label{V.21}
n_1(\bfr,\omega) = \! \int \! d^3r' \chi_s(\bfr,\bfr',\omega) \!\!\int \! d^3r'' f_{\rm Hxc}(\bfr',\bfr'',\omega)n_1(\bfr'',\omega)
\end{equation}
where we define the combined Hartree-xc kernel as $f_{\rm Hxc}(\bfr,\bfr',\omega)=|\bfr - \bfr'|^{-1}
+ f_{\rm xc}(\bfr,\bfr',\omega)$. Equation (\ref{V.21}) has the trivial solution $n_1=0$ for all frequencies $\omega$,
but at certain special frequencies $\Omega$ there are also nontrivial solutions where the density response is finite
and self-sustained, despite the fact that there is no external perturbation.
These frequencies correspond to the excitation energies of the system, and $n(\bfr,\Omega)$ is the profile of
the associated electronic eigenmode.

To illustrate how this works, consider the simple case of two electrons in a two-level system with Kohn-Sham orbitals
$\varphi_1(\bfr)$ and $\varphi_2(\bfr)$, assumed to be real. Each level is two-fold degenerate, and the lower level is doubly
occupied. Dropping the infinitesimal $i\eta$,
the Kohn-Sham response function (\ref{V.15}) then simplifies to
\begin{equation}\label{V.22}
\chi_s(\bfr,\bfr',\omega) = \frac{4\omega_{21}}{\omega^2 - \omega_{21}^2} \:\varphi_1(\bfr) \varphi_2(\bfr) \varphi_1(\bfr') \varphi_2(\bfr').
\end{equation}
We substitute this into Eq. (\ref{V.21}), and after a few simple manipulations we find the condition
\begin{equation} \label{V.23}
\omega^2 = \omega_{21}^2 + 4 \omega_{21} K(\omega)\:,
\end{equation}
where
\begin{equation} \label{V.24}
K(\omega) = \int \! d^3r \! \int \! d^3r' \varphi_1(\bfr) \varphi_2(\bfr) f_{\rm Hxc}(\bfr,\bfr',\omega) \varphi_1(\bfr')\varphi_2(\bfr').
\end{equation}
It is a simple exercise to repeat the above example using the spin-dependent response formalism. Assuming that the ground state
is not spin polarized (i.e., the spin-up and spin-down orbitals are the same), one finds the following
solutions for the eigenmodes:
\begin{equation} \label{V.25}
\omega^2_{\pm} = \omega_{21}^2 + 2 \omega_{21} [K_{\sigma\sigma}(\omega) \pm K_{\sigma\bar\sigma}(\omega)].
\end{equation}
The plus sign represents a singlet excitation, and the minus sign represents a triplet excitation.

The simple examples for two-level systems are instructive, but in practice turn out not to be quantitatively accurate
\cite{Petersilka1996,Vasiliev1999,Appel2003}. The eigenmodes can be calculated, in principle exactly, using
the so-called Casida equation \cite{Casida1995}:
\begin{equation} \label{V.26}
\left( \begin{array}{cc} {\bf A} & {\bf K} \\ {\bf K} & {\bf A} \end{array} \right)
\left( \begin{array}{c} {\bf X} \\ {\bf Y} \end{array} \right)
= \Omega
\left( \begin{array}{cc} -{\bf 1} & {\bf 0} \\ {\bf 0} & {\bf 1} \end{array} \right)
\left( \begin{array}{c} {\bf X} \\ {\bf Y} \end{array} \right),
\end{equation}
where the matrix elements of $\bf A$ and $\bf K$ are given by
\begin{eqnarray} \label{V.27}
A_{ia \sigma,i'a'\sigma'}(\omega) &=& \delta _{ii'} \delta_{aa'} \delta_{\sigma\sigma'} \omega_{ai \sigma} +
K_{ia \sigma,i'a'\sigma'}(\omega)
\\
K_{ia \sigma,i'a'\sigma'}(\omega) &=&\int d^3r \int d^3r' \varphi_{i\sigma}^*(\bfr) \varphi_{a\sigma}(\bfr)
\nonumber\\
&\times& f_{\rm Hxc\sigma\sigma'}(\bfr,\bfr',\omega) \varphi_{i'\sigma'}(\bfr')\varphi_{a' \sigma'}^*(\bfr') \label{V.28}
\hspace{5mm}
\end{eqnarray}
and $i,i'$ and $a,a'$ run over occupied and unoccupied Kohn-Sham orbitals, respectively. A detailed derivation of
Eq. (\ref{V.26}) can be found in Ref. \cite{Ullrich2012}.

If one assumes that
the Kohn-Sham orbitals are real and that the xc kernel is frequency-independent (more about this assumption in section \ref{subsec:VI.D}),
it is possible to recast the Casida equation into the following form:
\begin{eqnarray} \label{V.29}
\lefteqn{\hspace{-1cm}
\sum_{i'a'\sigma'} \Big[ \delta_{ii'} \delta_{aa'} \delta_{\sigma\sigma'}(\omega^2_{ia\sigma}-\Omega^2)
}\nonumber\\
&&
{}+ 2\sqrt{\omega_{ia\sigma}\omega_{i'a'\sigma'}} K_{ia\sigma,i'a'\sigma'}\Big]Z_{i'a'\sigma'}
=0 \:.
\end{eqnarray}
This equation can be viewed as the TDDFT counterpart of the eigenvalue equation (\ref{V.17}) for classical small oscillations.
Hence, Eq. (\ref{V.29}) yields the excitation energies and eigenmodes of the given system.

Eq. \parref{V.26} mixes excitations and de-excitations ($\mathbf{X}$ and $\mathbf{Y}$, respectively).
One may simplify Eq. \parref{V.26} by setting the off-diagonal $\mathbf{K}$ matrix to zero,
which decouples excitations and de-excitations. This so-called Tamm-Dancoff approximation (TDA) is
valid if the excitation frequencies are not close to zero, which is the case for molecules, semiconductors,
and insulators. The TDA often helps to compensate for deficiencies that arise because the xc functionals are
not exactly known and have to be approximated; the TDA can therefore be preferable
over the full calculation (in the sense of getting qualitatively correct results) in certain situations
(e.g. triplet instabilities \cite{Casida2000}, conical intersections \cite{Tapavicza2008}, and excitons \cite{Yang2012b}).

\subsection{Charge-transfer excitations} \label{subsec:VI.C}

An important class of excitations are those in which charge physically moves from one region (the donor)
to a second region (the acceptor) which is spatially separated from the first. Such processes can occur in a wide range of
systems, such as in complexes of two or more molecules, or between different functional groups within the same molecule.
Unfortunately, the standard approximations in TDDFT fail for charge-transfer excitations \cite{Dreuw2004,Hieringer2006,Autschbach2009}.

Consider the case where the donor and acceptor subsystems are separated by a large distance $R$.
The minimum energy required to remove an electron from the donor is given by the donor's ionization potential $I_d$.
When the electron attaches to the acceptor, some of that energy is regained via the acceptor's electron affinity $A_a$.
Once the electron has moved from donor to acceptor the two systems feel the electrostatic interaction energy $-1/R$ of the
induced electron--hole pair. The exact charge-transfer energy is therefore
\begin{equation} \label{9.CTexact}
\Omega_{ct}^{\rm exact} = I_d - A_a - \frac{1}{R} \:.
\end{equation}
Now let us compare this with TDDFT. To make our point it is sufficient to consider the two-level approximation, Eq. (\ref{V.23}).
After linearization, we obtain
\begin{eqnarray} \label{9.CTSPA}
\Omega_{ct} &=& \varepsilon^a_L - \varepsilon^d_H
+ 2\int \! d^3r \!\int \! d^3r' \: \varphi^a_L(\bfr) \varphi^d_H(\bfr)
\nonumber\\
&&{}\times f_{\rm Hxc}(\bfr,\bfr',\omega)  \varphi^a_L(\bfr') \varphi^d_H(\bfr') \:,
\end{eqnarray}
where $\varphi^d_H(\bfr)$ is the highest occupied donor orbital and $ \varphi^a_L(\bfr)$ is the lowest unoccupied acceptor orbital,
which have exponentially vanishing overlap in the limit of large separation. Hence,
the double integral in Eq. (\ref{9.CTSPA}) becomes zero (assuming that the xc kernel remains finite, which is certainly the case
for all standard approximations), and TDDFT simply collapses to the difference between the bare Kohn--Sham eigenvalues,
\begin{equation} \label{9.CTSPAlocal}
\Omega_{ct} \longrightarrow \varepsilon^a_L - \varepsilon^d_H \:.
\end{equation}
This explains why TDDFT often drastically underestimates charge-transfer
excitations when conventional xc functionals are used.
Hybrid xc functionals \cite{Zhao2006a,Zhao2006b}, in particular the range-separated hybrids of Section \ref{subsec:II.E},
offer a solution to this problem, and have been successfully used to describe charge-transfer excitations
in a variety of systems \cite{Tawada2004,Stein2009,Kronik2012}.

\subsection{Beyond the adiabatic approximation} \label{subsec:VI.D}

The exact excitation spectrum of a physical system is determined by the poles of the full response function $\chi$, Eq. (\ref{V.D.5}).
All of the excitation energies $\Omega_n$ of the many-body system are, in principle, obtained by solving
the Casida equation (\ref{V.26}). But it is found that within the adiabatic approximation for $f_{\rm xc}$, some of the excitations
are missing \cite{Tozer2000,Hsu2001,Cave2004}! The missing excitations turn out to be those that have the
character of double (or multiple) excitations, i.e., the associated many-body excited states, if expanded
in a basis of Kohn-Sham Slater determinants, contain dominant contributions of doubly excited configurations.

The Kohn-Sham noninteracting response function $\chi_s$ (\ref{V.15}) has poles at the Kohn--Sham single
excitations. Compared with the many-body response function (\ref{V.D.5}), $\chi_s$ has fewer poles,
since a noninteracting system cannot have double and multiple excitations in linear response.
Solving the Casida equation in a finite basis and using the
adiabatic approximation for $f_{\rm xc}$, as is done in practice, will not change the number of poles,
but just shift them. To obtain double excitations, a frequency-dependent $f_{\rm xc}(\omega)$ is needed
which will generate additional solutions, since the Casida equation then becomes a nonlinear eigenvalue problem.

Thus, we can say the following about the adiabatic approximation in TDDFT:
\begin{itemize}
\item The adiabatic approximation works well for those excitations of the physical system for
which a correspondence to a single excitation in the Kohn-Sham system exists. The Casida equation
then shifts the Kohn-Sham excitations towards the true single excitations.

\item The frequency dependence of $f_{\rm xc}$ must kick in for those excitations of the physical
system that are missing in the Kohn-Sham system, namely, double or multiple excitations.
\end{itemize}

Several nonadiabatic TDDFT approaches for the description of molecular double excitations have been
explored in the literature. One of them is known as dressed TDDFT  \cite{Maitra2004}, where
a frequency-dependent xc kernel is explicitly constructed within a small subspace.
Other nonadiabatic approaches are based on many-body theory \cite{Gritsenko2009,Romaniello2009,Sangalli2011,Sakkinen2012}.
However, none of these approaches is sufficiently straightforward to be part of
mainstream TDDFT.

\subsection{Periodic systems and long-range behavior} \label{subsec:VI.E}

As seen from Eqs. \parref{V.27} and \parref{V.28}, the Casida equation is expressed in the space
spanned by one-particle Kohn-Sham transitions \cite{Furche2001}. Real-space
kernels are suitable for calculations of finite systems such as atoms and molecules. For periodic systems
like solids, the momentum space representation of the Hartree-xc kernel is more convenient. In Section \ref{subsec:VII.D}
we will use this approach to describe the optical properties of insulating solids.

The real space
representation of the kernel is related to the momentum space representation as
\begin{multline}
f_{{\rm Hxc}\sigma\sigma'}(\bfr,\bfr',\omega)=\frac{1}{V}\sum_{\vect{q}\in{\rm FBZ}}\sum_{\vect{G},\vect{G}'}e^{i(\vect{q}+\vect{G})\cdot\bfr}\\
\times f_{{\rm Hxc}\sigma\sigma'}(\vect{q},\vect{G},\vect{G}',\omega)e^{-i(\vect{q}+\vect{G}')\cdot\bfr'},
\label{eqn:application:longrange:fxcqspace}
\end{multline}
where $\vect{G}$, $\vect{G}'$ are reciprocal
lattice vectors. With Eq. \parref{eqn:application:longrange:fxcqspace}, the Hartree-xc kernel in
transition space, Eq. \parref{V.28}, becomes
\begin{multline}
K_{ia\sigma,i'a'\sigma'}=\frac{1}{V}\sum_{\vect{q}\in{\rm FBZ}}\sum_{\vect{G},\vect{G}'}
\langle i\vect{k}_i\sigma | e^{i(\vect{q}+\vect{G})\cdot\bfr}| a\vect{k}_a\sigma \rangle
\\
\times f_{{\rm Hxc}\sigma\sigma'}(\vect{q},\vect{G},\vect{G}')
\langle a'\vect{k}_{a'}\sigma' | e^{-i(\vect{q}+\vect{G}')\cdot\bfr'}| i'\vect{k}_{i'}\sigma' \rangle \\
\times\delta_{\vect{k}_a-\vect{k}_i+\vect{q},\vect{G}_0}\delta_{\vect{k}_{a'}-\vect{k}_{i'}+\vect{q},\vect{G}'_0},
\label{eqn:application:longrange:Kreci}
\end{multline}
with the matrix elements defined as
\begin{equation}
\langle i\vect{k}_i\sigma | e^{i(\vect{q}+\vect{G})\cdot\bfr} | a\vect{k}_a\sigma \rangle   \equiv\int \! d^3r
\phi_{i\vect{k}_i\sigma}^*(\bfr)e^{i(\vect{q}+\vect{G})\cdot\bfr}\phi_{a\vect{k}_a\sigma}(\bfr),
\end{equation}
where $\vect{k}$'s are the Bloch wavevectors of the corresponding wavefunctions, and ${\bf G}_0$, ${\bf G}'_0$
can be any reciprocal lattice vector. The Kronecker-$\delta$s in Eq. \parref{eqn:application:longrange:Kreci}
are a consequence of Bloch's theorem.

The Hartree part of $f_{\rm Hxc}$ can be shown to be largely irrelevant for the optical properties of insulators close to the gap \cite{Onida2002};
we therefore focus on the xc part in the following. For $\vect{G}=\vect{G}'=0$ (the so-called {\em head}
of $f_{\rm xc}$) in the important limit of $\vect{q}\to0$, which corresponds to infinite range in real space, both matrix elements in
Eq. \parref{eqn:application:longrange:Kreci} behave as $O(q^1)$. All the local and semilocal
xc kernels (derived from LDA and GGA in the adiabatic approximation) have finite values for the head.
Since the two matrix elements in Eq. \parref{eqn:application:longrange:Kreci} together
vanish as $O(q^2)$, the head contribution to the sum of Eq. \parref{eqn:application:longrange:Kreci}
is zero for all (semi)local kernels. For these kernels, all changes to the Kohn-Sham spectrum
come from the {\em body} of $f_{\rm xc}$ (where $\vect{G}\ne 0,\,\vect{G}'\ne 0$).

Gonze {\em et al.} \cite{Gonze1995,Ghosez1997} pointed out that the head of $f_{\rm xc}$ has to
diverge as $q^{-2}$ for $q\to0$ to correctly describe the polarization of periodic insulators.
With the $q^{-2}$ divergence, the head of $f_{\rm xc}$ contributes in the sum of
Eq. \parref{eqn:application:longrange:Kreci}, dominating
the other parts of $f_{\rm xc}$ [wings ($\vect{G}=0,\vect{G}'\ne 0$ or vice versa) and body].
Local and semilocal xc kernels do not have this long-range
behavior, and there is no obvious and consistent way of modifying them
to include the long-rangedness.

The long-range behavior of the xc kernel is unimportant
for low-lying excitations in finite systems such as atoms and molecules,
which means that local and semilocal xc kernels will work reasonably well. However,
for extended and periodic systems it is crucial to have xc kernels with the proper long-range behavior
to obtain correct optical spectra \cite{Onida2002,Botti2004}. We will discuss this further in Section \ref{subsec:VII.D}.


\section{Applications in linear response} \label{sec:VII}

Linear-response TDDFT has been implemented in many computer codes in quantum chemistry
and materials science. In this Section we will give an overview of some of the most important
areas of application.

\subsection{Standard approximations for the xc kernel} \label{subsec:VII.A}

To carry out a TDDFT calculation in the linear response formalism, one must know the xc kernel.
The simplest thing to do is to use the random-phase approximation (RPA), where the xc kernel is set to zero:
\begin{equation}
f_{\rm xc}^{\rm RPA}(\bfr,\bfr',\omega)=0.
\end{equation}
This seemingly trivial kernel originates from many-body theory, where one sums up all the `bubble'
type diagrams \cite{GiulianiVignale}. Though the form is similar to time-dependent Hartree, TDDFT
RPA is fundamentally different due to the use of the Kohn-Sham system.
The RPA kernel has seen applications for molecules and is known to produce reasonably
good results. For insulating solids, the RPA spectra are missing
important features such as excitonic effects (see below).

The proper way to obtain $f_{\rm xc}$ is via Eq. \parref{V.12}: first approximate the time-dependent xc potential,
calculate $f_{\rm xc}(\bfr,t,\bfr',t')$ by taking the functional derivative,
and then get the frequency-dependent kernel $f_{\rm xc}(\bfr,\bfr',\omega)$ via Fourier transform.
However, these steps are rarely carried out in practice, since most of the xc kernels in use are adiabatic kernels.
Recall the adiabatic approximation for the xc potential, Eq. (\ref{V.1a}), which uses the ground-state
functional and evaluates it at the time-dependent density. The adiabatic approximation for the xc kernel is
\begin{equation}
f_{\rm xc}^{\rm A}(\bfr,\bfr') = \frac{\delta v_{\rm xc}^{\rm gs}[n_0](\bfr)}{\delta n_0(\bfr')}
= \frac{\delta^2 E_{\rm xc}[n_0]}{\delta n_0(\bfr) \delta n_0(\bfr')} \:,
\end{equation}
which is frequency-independent.

An important example is the ALDA xc kernel:
\begin{equation}
f_{\rm xc}^{\rm ALDA}(\bfr,\bfr') = \left. \frac{d^2 e_{\rm xc}^h(\bar n)}{d \bar n ^2}\right|_{\bar n = n_0(\bfr)}
\delta(\bfr-\bfr') \:,
\label{eqn:ALDAxc}
\end{equation}
whose exchange part is explicitly given by
\begin{equation}
f_{\rm x}^{\rm ALDA}(\bfr,\bfr')=-[9\pi n_0^2(\bfr)]^{-1/3}\delta(\bfr-\bfr'),
\end{equation}
and the correlation part can be obtained by using Eq. \parref{eqn:ALDAxc} on
any of the interpolations of $e_{\rm c}^{\rm LDA}$\cite{Vosko1980,Perdew1981,Perdew1992}.

This xc kernel is not only frequency-independent, it is also local.
One can derive adiabatic-GGA (AGGA) kernels in a similar fashion, starting from any of the standard GGA functionals
such as those discussed in Section \ref{subsec:II.E}. Adiabatic hybrid kernels, most notably B3LYP, are
very widely used and have contributed much to the success of TDDFT in quantum chemistry.

\subsection{Molecular excitations} \label{subsec:VII.B}

\begin{table}[tb]
{
\caption{Low-lying excitation energies (in eV) of the benzene molecule $(\rm C_6 H_6)$ calculated with TDDFT
using various xc functionals with the basis set 6-31++G(3df,3pd), and geometry optimized using the respective
functionals with the same basis \protect\cite{Tao2008}. CASPT2 reference results (Ref), TDHF, and experimental results from Packer {\em et al.}\
\protect\cite{Packer1996}. The mean average error (MAE) for TDHF excludes the lowest ($^3B_{1u}$) triplet transition,
which comes out unstable. \label{table3}}
}
{\begin{tabular}{lcccccccc}
\hline
 & & \\[-8pt]
 & LDA & PBE & TPSS & PBE0 & B3LYP & Ref & TDHF & Exp \\[2pt]
\hline
 & & \\[-6pt]
$^3B_{1u}$  & 4.47 & 3.98 & 3.84 & 3.68 & 3.84 & 3.89 &   ---  & 3.94 \\[3pt]
$^3E_{1u}$  & 4.82 & 4.61 & 4.67 & 4.75 & 4.72 & 4.49 & 4.70 & 4.76 \\[3pt]
$^1B_{2u}$  & 5.33 & 5.22 & 5.32 & 5.52 & 5.41 & 4.84 & 5.82 & 4.90 \\[3pt]
$^3B_{2u}$  & 5.05 & 4.89 & 4.98 & 5.12 & 5.07 & 5.49 & 5.57 & 5.60 \\[3pt]
$^1B_{1u}$  & 6.07 & 5.94 & 6.00 & 6.18 & 6.05 & 6.30 & 5.88 & 6.20 \\[3pt]
$^1E_{1g}$  & 6.12 & 5.89 & 5.99 & 6.38 & 6.11 & 6.38 & 6.54 & 6.33 \\[3pt]
$^1A_{2u}$  & 6.70 & 6.43 & 6.50 & 6.90 & 6.62 & 6.86 & 6.94 & 6.93 \\[3pt]
$^1E_{2u}$  & 6.71 & 6.44 & 6.50 & 6.95 & 6.65 & 6.91 & 7.11 & 6.95 \\[4mm]
MAE     & 0.30 & 0.37 & 0.33 & 0.18 & 0.27 & 0.09  & $0.26^*$\\
\hline
\end{tabular}
}
\end{table}

As an example, let us consider the benzene molecule.
Table \ref{table3} shows eight low-lying singlet and triplet excitation energies of
benzene, calculated with various xc functionals \cite{Tao2008}. As an overall measure
of the accuracy of the calculations, the mean absolute error (MAE) was also calculated for each functional.
Based on this measure, the nonhybrid xc functionals (LSD, PBE, and TPSS) perform at about the same level,
with an MAE of 0.3-0.4 eV. The hybrid functionals (PBE0 and B3LYP) used in this study perform
somewhat better, with an MAE ranging from 0.18 to 0.27 eV. As we will see in the following examples, these
findings are quite typical.

\begin{figure}
\centering
\includegraphics[width=0.8\linewidth]{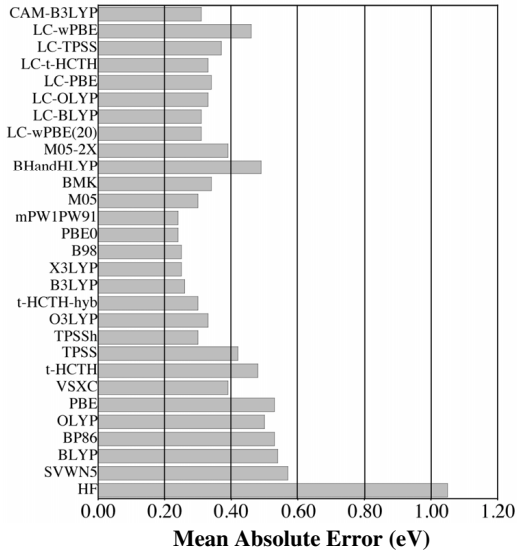}
\caption{Mean absolute error for the lowest vertical excitation energies of a test set of 28 medium-sized organic
molecules (103 excited states). Reproduced with permission from ACS from Ref. \cite{Jacquemin2009}. \copyright 2009.}
\label{fig9}
\end{figure}

Figure \ref{fig9} shows the MAE for 28 xc functionals and for HF, obtained by
calculating 103 low-lying vertical excitation energies for a test set of 28 medium-sized organic
molecules \cite{Jacquemin2009}, compared against accurate theoretical benchmarks.
The Kohn--Sham ground states were obtained with the
same xc functionals that were used, in the adiabatic approximation, for the TDDFT calculations.
Identical molecular geometries were used for each xc functional.

TDHF gives very large errors (over 1 eV), almost always overestimating the transition energies;
any TDDFT calculation reduces the error by at least a half.
Among the xc functionals, we can distinguish between pure density functionals (LDA and GGA), meta-GGAs,
hybrid GGAs, and long-range-corrected hybrids (the first eight functionals in Fig. \ref{fig9}).
The LDA and GGAs all give an MAE of order 0.5 eV. Meta-GGAs (VSXC and TPSS) give better
agreement (about 0.4 eV). But the best choice are clearly the hybrid GGAs (B3LYP, X3LYP, B98,
mPW1PW91, and PBE0). In this case, the MAE is reduced to less than 0.25 eV. Similar findings were also reported in a more recent
benchmark study \cite{Leang2012}.

The long-range-corrected (LC) hybrids such as CAM-B3LYP
give a slightly higher error, owing to a general overestimation of the transition
energies. This is mainly due to the choice of the test set, in which charge-transfer excitations are
not significantly represented. The advantage of long-range-corrected
hybrids emerges for such excitations in larger molecules.

\begin{figure}
\centering
\includegraphics[width=\linewidth]{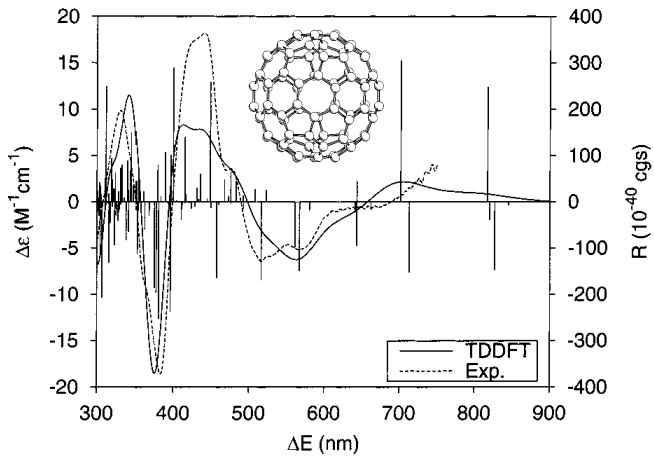}
\caption{Circular dichroism spectrum of $D_2$-$\rm C_{84}$, comparing TDDFT with experiment.
($\varepsilon$: molar decadic absorption coefficient; $R$: rotatory strength;
$\Delta E$: excitation energy).
Reproduced with permission from ACS from Ref. \cite{Furche2002}. \copyright 2002.}
\label{fig10}
\end{figure}

\begin{figure}
\centering
\includegraphics[width=0.8\linewidth]{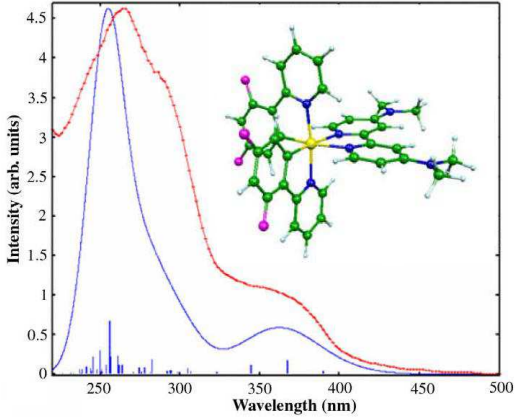}
\caption{Calculated (blue line) and experimental (red line) absorption spectra of a
Iridium(III) cyclometallated complex.
Blue vertical lines correspond to the unbroadened oscillator strength of the calculated singlet-singlet transitions.
Reproduced with permission from Elsevier from Ref. \cite{DeAngelis2009}. \copyright 2009.}
\label{fig11}
\end{figure}

As these examples illustrate, TDDFT offers an excellent compromise between computational efficiency
and accuracy. TDDFT scales as $N^2$ to $N^3$, depending on the implementation;
wave-function-based methods of comparable accuracy scale at least one or two orders of
magnitude worse. The current limit of high-end wave-function-based methods is about 50 atoms
\cite{Grimme2004,Dreuw2005,Jacquemin2009}. By contrast, TDDFT allows the treatment of molecules containing hundreds of
atoms. Examples of medium-sized systems are shown in Figs. \ref{fig10} and \ref{fig11}.

Figure \ref{fig10} shows the circular dichroism spectrum of a large chiral fullerene molecule. TDDFT was able
to resolve a debate regarding the molecular configuration of this system \cite{Furche2002}.
Figure \ref{fig11} shows the absorption spectrum of an Iridium(III) cyclometallated complex \cite{DeAngelis2009}.

\subsection{Potential-energy surfaces} \label{subsec:VII.C}

Consider a system with $N_e$ electrons and $N_n$ nuclei, with nuclear masses $M_j$ and charges $Z_j$, where $j=1,\ldots,N_n$.
Formally, all electrons and all nuclei are quantum
mechanical particles, forming an interacting $N_e+N_n$-body system. For instance, the $H_2$ molecule
depends on the coordinates of the two electrons, $\bfr_1$ and $\bfr_2$, and on the coordinates of the
two protons, $\bfR_1$ and $\bfR_2$: hence, it is a four-body problem.

We denote the sets of electronic and nuclear spatial coordinates by $\mbbr \equiv \{\bfr_1,\ldots,\bfr_{N_e}\}$ and
$\mbbR \equiv \{\bfR_1,\ldots,\bfR_{N_n}\}$, respectively.
The many-body eigenstates of the system are a function of the
two sets of coordinates, $\mbbPsi_j(\mbbr,\mbbR)$, and
obey the following many-body Schr\"odinger equation
\begin{equation} \label{17.TDSE}
\hat{H}(\mbbr,\mbbR) \mbbPsi_i(\mbbr,\mbbR,t)
= E_i \mbbPsi_i(\mbbr,\mbbR,t)
\:.
\end{equation}
In the absence of any external potentials, the
Hamiltonian of the coupled electron-nuclear system is given by
\begin{eqnarray} \label{17.Htot}
\hat{H}(\mbbr,\mbbR) &=& -\sum_{j=1}^{N_e} \frac{\nabla_{\bfr_j}^2}{2}
+ \frac{1}{2}\sum_{{j,k}\atop{j\ne k}}^{N_e} \frac{1}{|\bfr_j - \bfr_k|}
-\sum_{j=1}^{N_n}\frac{\nabla_{\bfR_j}^2}{2M_j}
\nonumber\\
&+&
\frac{1}{2}\sum_{{j,k}\atop{j\ne k}}^{N_n} \frac{Z_j Z_k}{|\bfR_j - \bfR_k|} -
\sum_{j=1}^{N_e}\sum_{k=1}^{N_n} \frac{Z_k}{|\bfr_j - \bfR_k|}
\nonumber\\
&\equiv& \hat{T}_e + \hat{W}_{ee} + \hat{T}_n  + \hat{W}_{nn} + \hat{W}_{en} \:.
\end{eqnarray}
As can be seen, $\hat{H}(\mbbr,\mbbR)$ is the sum of an electronic Hamiltonian containing
kinetic energy and electron-electron interaction,
$\hat{T}_e + \hat{W}_{ee}$, a similar nuclear Hamiltonian $\hat{T}_n  + \hat{W}_{nn}$,
and an electron-nuclear coupling term $\hat{W}_{en}$.

The full coupled electron-nuclear many-body problem is too difficult to solve in general;
one usually works in the Born-Oppenheimer (BO) approximation to obtain the structure of
molecules and solids.
The central idea of the BO approximation is that because of the large
 difference between the electronic and nuclear masses (the proton
is 1836 times more massive than the electron), the two sets of degrees of freedom are essentially decoupled.

The BO Hamiltonian is defined as the full Hamiltonian (\ref{17.Htot}) minus the
nuclear kinetic-energy term:
\begin{eqnarray} \label{17.HBO}
\hat{H}_{\rm BO}(\mbbr,\mbbR) &=& -\sum_{j=1}^{N_e}\frac{\nabla_{\bfr_j}^2}{2}
+ \frac{1}{2}\sum_{{j,k}\atop{j\ne k}}^{N_e} \frac{1}{|\bfr_j - \bfr_k|}
\nonumber\\
&+&
\frac{1}{2}\sum_{{j,k}\atop{j\ne k}}^{N_n} \frac{Z_j Z_k}{|\bfR_j - \bfR_k|}
- \sum_{j=1}^{N_e}\sum_{k=1}^{N_n} \frac{Z_k}{|\bfr_j - \bfR_k|} \:.
\end{eqnarray}
This Hamiltonian depends parametrically on the nuclear coordinates: this means that the nuclear positions $\bfR_1,\ldots,\bfR_{N_n}$
are just treated as a set of given numbers, indicating a given nuclear configuration; they are no longer quantum mechanical operators.
For each configuration one solves the Schr\"odinger equation
\begin{equation} \label{17.SEBO}
\hat{H}_{\rm BO}(\mbbr,\mbbR) \Psi_j(\mbbr,\mbbR) =
E_j (\mbbR) \Psi_j(\mbbr,\mbbR) \:.
\end{equation}
The energy eigenvalues $E_j (\mbbR)$ define the landscape of potential-energy surfaces, whose dimensionality
depends on the degrees of freedom of the molecule. Thus, for a diatomic molecule, $E_j (\mbbR)$ can be represented simply by a curve
as a function of the internuclear distance, whereas for $N_n\ge 3$ it is a function of $3N_n-6$ coordinates
($3N_n-5$ for linear molecules) and should therefore more appropriately
be called a ``hypersurface'';  the potential-energy surface is a 2D section through this higher-dimensional space. In common usage, however,
the distinction between a surface and a hypersurface is usually not made.

The ground-state potential-energy surface $E_0 (\mbbR)$ is of particular interest because its
minimum defines the molecular equilibrium position. However, excited-state potential-energy surfaces are important too, and
play a crucial role in chemical reactions, photochemical processes, and in spectroscopy.

All potential-energy surfaces following from Eq. \parref{17.SEBO} are called
{\em adiabatic}, indicating a complete decoupling of electronic and nuclear degrees of freedom.
The calculation of adiabatic potential-energy surfaces is one of the key tasks of computational chemistry.
The lowest potential-energy surface can be obtained exactly, in principle,
using ground-state DFT; for excited-state potential-energy surfaces, forces, and vibrational frequencies,
the appropriate method is TDDFT \cite{Furche2002,Liu2011}.

\begin{figure}
\begin{center}
\includegraphics[angle=0, width=8.65cm]{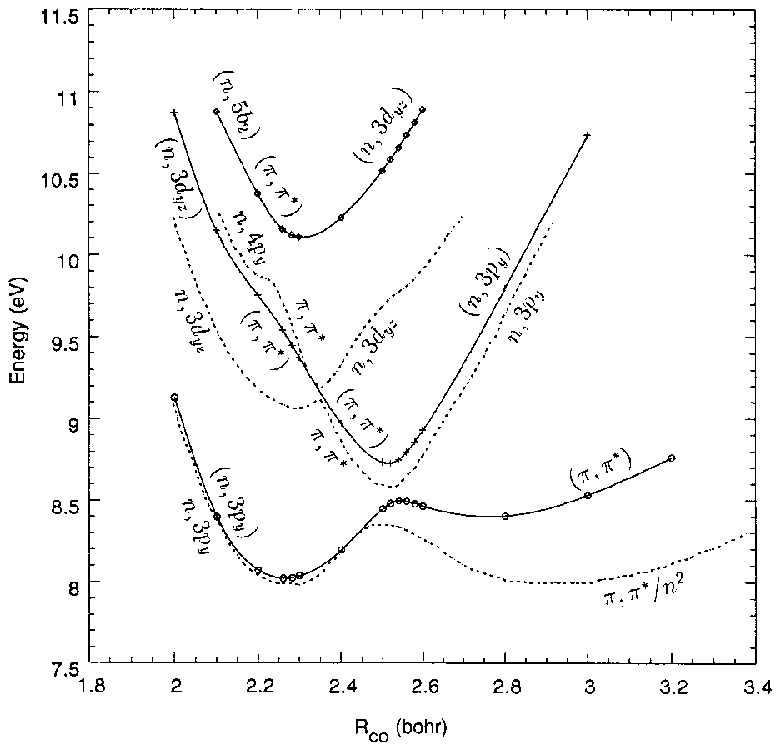}
\end{center}
\caption{$1A_1$ CO-stretch potential-energy curves of planar formaldehyde ($\rm CH_2O$). Full lines: TDDFT. Dashed lines: multireference
doubles CI. Reproduced with permission from Wiley from \protect\cite{Casida1998b}. \copyright 1998.
}
\label{fig17.2}
\end{figure}

Figure \ref{fig17.2} shows the $1A_1$ manifold of the CO-stretch potential-energy curves of planar formaldehyde \cite{Casida1998b}.
These are excited states, several eV above the ground-state potential-energy curve (whose minimum is set at 0 eV). The dashed lines
are results from a multireference doubles CI benchmark calculation; the full lines were obtained with TDDFT, using the ALDA
with an asymptotic correction. An xc functional with the correct asymptotics is important here because
these are high-lying (Rydberg) excitations.

A prominent feature in Fig.\ \ref{fig17.2} is the avoided crossing between the states labeled $(\pi,\pi^*)$
and $(n,3p_y)$. TDDFT reproduces this
avoided crossing qualitatively correctly, thanks to the configuration mixing of individual single-particle transitions
induced by the off-diagonal matrix elements $K_{ia\sigma,i'a'\sigma'}$ in the Casida equation (\ref{V.26}) \cite{Maitra2006a}.

The $(n,3p_y)$ curve is almost on top of the exact curve, at least for C--O distances before the avoided crossing. On the other hand,
the $(n,3d_{yz})$ curve comes out about 1 eV too high, primarily owing to limitations of the xc functional used in this calculation.

There are many  TDDFT studies in organic and inorganic photochemistry calculating
excited-state potential-energy surfaces  \cite{Sobolewski1999,Wanko2004,Tachikawa2004,Cordova2007,Tsai2010}.
The performance of TDDFT depends strongly on the xc functional
used (choosing appropriate basis sets is another important factor). Complications can
arise for potential-energy surfaces
associated with excitations that have a long-range, charge-transfer character \cite{Wiggins2009,Plotner2010}.
In that case, local or semilocal xc functionals will fail, and one needs to use xc functionals
with the correct long-range behavior, see Section \ref{subsec:VI.C}.

Another source of complications are situations in which the ground state has an intrinsically multiconfigurational
character. This can lead to circumstances in which two potential-energy surfaces become degenerate and touch each other,
which gives rise to so-called conical intersections. The name reflects the topology in the vicinity of the point
of degeneracy, which looks like an inverted cone balancing on the tip of another cone.
TDDFT has serious problems with conical intersections \cite{Levine2006,Tapavicza2008,Kaduk2010}:
it typically produces the wrong topology in the vicinity of the intersection point.
These difficulties have a lot to do with the problems of TDDFT to describe double excitations:
an explicitly frequency-dependent xc kernel $f_{\rm xc}(\omega)$ is required for a proper description
of conical intersections \cite{Maitra2006a}.

\subsection{Optical properties of solids} \label{subsec:VII.D}

At present, the majority of applications of TDDFT are in the area of computational (bio)chemistry.
However, applications in solid-state physics and materials science are emerging at a rapid rate.
In this Section we will highlight some of the most important issues for TDDFT in solids:
the band-gap problem, excitons in insulators, and plasmons in metals.

\subsubsection{The band gap versus the optical gap}

The fundamental band gap $E_g$ is a key quantity that characterizes insulating materials.
It  is defined as follows:
\begin{equation} \label{2.gapIA}
E_g(N) = I(N)-A(N),
\end{equation}
where $I(N)$ and $A(N)$ are the ionization potential and the electron affinity of the $N$-electron system,
see Eqs. (\ref{IN}) and (\ref{AN}). Hence, we obtain
\begin{equation}\label{Eg}
E_g(N) = \varepsilon_{N+1}(N+1) - \varepsilon_{N}(N) \:.
\end{equation}
It is important to note that the right-hand side of Eq. (\ref{Eg}) contains the highest occupied Kohn--Sham
eigenvalues of two different systems, namely with $N$ and with $N+1$ electrons. In a macroscopic solid with $10^{23}$ electrons,
it would of course be impossible to calculate the band gap according to this definition.

The band gap in the noninteracting Kohn--Sham system, also known as the Kohn--Sham gap, is defined as
\begin{equation}\label{EKS}
E_{g,s}(N) = \varepsilon_{N+1}(N) - \varepsilon_N(N) \:.
\end{equation}
In contrast with the interacting gap $E_g$, the Kohn--Sham gap $E_{g,s}$ is simply the difference between the highest occupied and
lowest unoccupied single-particle levels in the {\em same} $N$-particle system. This quantity is what is usually taken
as the band gap in standard DFT band-structure calculations. We can relate the two gaps by
\begin{equation} \label{2.Deltaxcdef}
E_g = E_{g,s} + \Delta_{\rm xc},
\end{equation}
which defines $\Delta_{\rm xc}$ as a many-body correction to the Kohn--Sham gap. By making use of the previous relations,
we find $\Delta_{\rm xc} = \varepsilon_{N+1}(N+1) - \varepsilon_{N+1}(N)$.
It turns out that the many-body gap correction $\Delta_{\rm xc}$ can be related to a very fundamental property of
density functionals, known as derivative discontinuities \cite{Perdew1982,Perdew1983,Godby1986,Capelle2010}.

The so-called band-gap problem of DFT reflects the fact that in practice $E_{g,s}$ is often a poor
approximation to $E_g$, typically underestimating the exact band gap by as much as $50\%$.
The reason for this is twofold: commonly used approximate xc functionals (such as LDA and GGA) tend to underestimate the
{\em exact} Kohn-Sham gap $E_{g,s}$, and they do not yield any discontinuity correction $\Delta_{\rm xc}$.
An extreme example for the second failure are Mott insulators, which are typically predicted to be metallic by
DFT. This is no accident: in Mott insulators, the exact Kohn-Sham system is metallic (i.e., $E_{g,s}=0$)
so that $E_g=\Delta_{\rm xc}$. Clearly, standard xc functionals (where $\Delta_{\rm xc}$ vanishes) are unfit
to describe Mott insulators.

\begin{figure}
\begin{center}
\includegraphics[angle=0, width=7.5cm]{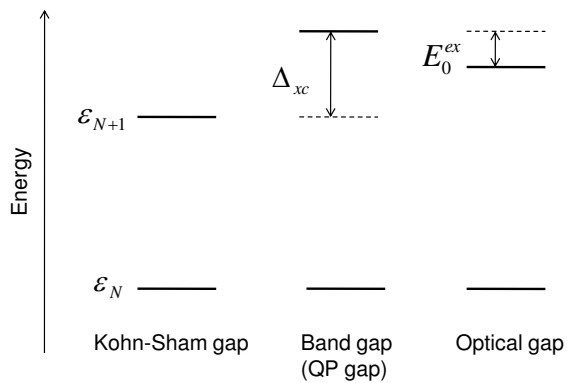}
\end{center}
\caption{Schematic illustration of the different types of gaps in DFT and TDDFT. The Kohn-Sham gap is defined as the difference
of the highest occupied and lowest unoccupied Kohn-Sham eigenvalues of the $N$-electron system, see Eq. (\ref{EKS}).
The fundamental band gap [or quasiparticle (QP) gap] is the Kohn-Sham gap plus the derivative discontinuity, see Eq. (\ref{2.Deltaxcdef}).
The optical gap is the band gap minus the lowest exciton binding energy $E_0^{\rm ex}$. The Kohn-Sham gap
can be viewed as an approximation for the optical gap.
}
\label{fig_gap}
\end{figure}

It is important to distinguish between the fundamental band gap
and the optical gap \cite{Kronik2012}. The band gap describes the energy that an electron
must have so that, when added to an $N$-electron system, the result is an $N+1$ electron
system in its ground state. The total charge of the system changes by $-1$ in
this process. By contrast, the optical gap describes the lowest neutral excitation of an $N$-electron system:
here, the number of electrons remains unchanged. The two gaps are schematically illustrated in Fig. \ref{fig_gap}
together with the Kohn-Sham gap.

The band gap of insulators can be accurately obtained from the so-called quasiparticle energies, which are defined as the
single-particle energies of a noninteracting system whose one-particle Green's function is the same as that of the real interacting
system (note how this is different from the definition of the Kohn-Sham system). In practice, this is
often done using the GW method \cite{Hedin1965,Aryasetiawan1998,Onida2002}. GW calculations are more demanding than
DFT, but they produce band structures of solids that agree very well with experiment. Generalized Kohn-Sham schemes \cite{Seidl1996,Kummel2008}
can also give good band gaps.

While the band gap can be measured using techniques in which electrons are added or removed from the system
(such as photoemission spectroscopy), the optical gap refers to the lowest neutral excitation.
The difference between quasiparticle band gap and optical gap is the
lowest exciton binding energy, $E_0^{\rm ex}$. Excitons can be viewed as bound electron-hole pairs, whose
bound states form a Rydberg series, analogous to the hydrogen atom \cite{Yang2012b}. The band gap is given by the asymptotic limit
of the excitonic Rydberg series \cite{Perdew2009} (at least for direct-gap insulators and semiconductors).

TDDFT can be used to calculate optical spectra of materials in principle exactly. In the case of
insulators and semiconductors, this means that it should, in principle, yield the correct optical gap, the
correct excitonic Rydberg series (if the material under study has one), and hence the correct band gap (obtained as the limit of
the excitonic Rydberg series). We will discuss in detail in the following section how optical spectra of insulators
and semiconductors are calculated with TDDFT in practice.

As always in TDDFT, the burden rests on the xc kernel. In the case of bulk insulators, $f_{\rm xc}$
needs to accomplish two things: it needs to ``open up'' the gap (i.e., compensate the fact that the Kohn-Sham gap
underestimates the band gap), and it needs to produce the electron-hole interaction that is responsible for
the formation of excitons. Formally, we can write this as follows \cite{Stubner2004}:
\begin{equation}
f_{\rm xc} = f_{\rm xc}^{\rm qp} + f_{\rm xc}^{\rm ex} \:.
\end{equation}
The xc kernel is written as the sum of a quasiparticle part $f_{\rm xc}^{\rm qp}$ (which opens up the gap)
and an excitonic part $f_{\rm xc}^{\rm ex}$ (which causes excitonic effects). The excitonic part turns out
to be easier to approximate than the quasiparticle part (see below). In fact, no suitable approximations to
$f_{\rm xc}^{\rm qp}$ exist at present. To a large extent this is  due to the fact that the quasiparticle part is
intrinsically nonadiabatic \cite{Gonze1999}: the frequency-dependence is essential to shift the Kohn-Sham gap,
and to produce an excitonic Rydberg series \cite{Yang2012b}. In view of this, one usually ignores
the quasiparticle part of $f_{\rm xc}$ and starts from a band structure in which the
gap has been corrected by other means (such as via GW, or with a simple scissor operator \cite{Levine1989}).

\subsubsection{Optical spectra of semiconductors and insulators}

\begin{figure}
\begin{center}
\includegraphics[angle=0, width=6.5cm]{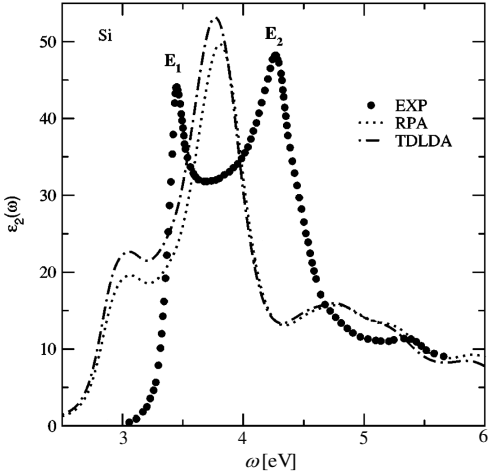}
\end{center}
\caption{Optical absorption spectrum of bulk Si. RPA and TDLDA fail to reproduce the optical gap and
the excitonic peak. Reproduced with permission from APS from \cite{Botti2004}. \copyright 2004.}
\label{fig:SiSpectrum}
\end{figure}

In the optical spectroscopy of solids, a central quantity is the complex index of refraction $\tilde n$, defined as
\cite{Yu2010}
\begin{equation}
\tilde n^2 = \epsilon_{\rm mac}(\omega) \:,
\end{equation}
where $\epsilon_{\rm mac}(\omega)$ is the macroscopic dielectric function. The imaginary part
of  $\epsilon_{\rm mac}(\omega)$ hence describes the photoabsorption of a solid, as illustrated
in Fig. \ref{fig:SiSpectrum} for the case of silicon.
To calculate the macroscopic dielectric function from first principles, we need to take a detour and
first calculate the {\em microscopic} dielectric matrix, $\epsilon({\bf q},{\bf G},{\bf G}',\omega)$,
where $\bf G$ and $\bf G'$ are reciprocal lattice vectors. The macroscopic dielectric function then follows as the limit \cite{Onida2002}
\begin{equation}
\epsilon_{\rm mac}(\omega)= \lim_{\bfq \to0}\frac{1}{\epsilon^{-1}(\bfq,\bfG=0,\bfG'=0,\omega)} \:.
\label{eqn:epsilonmac}
\end{equation}
In turn, the inverse dielectric function of a periodic system can be obtained from the response function as
\begin{equation}
\epsilon^{-1}(\bfq,\bfG,\bfG',\omega) = \delta_{\bfG \bfG'} + v_\bfG(\bfq) \chi(\bfq,\bfG,\bfG',\omega) \:,
\label{eqn:inverseepsilon}
\end{equation}
where $v_\bfG(\bfq) = 4\pi/|\bfq + \bfG|^2$. In TDDFT, the full response function is
expressed as
\begin{multline}
\chi(\vect{q},\vect{G},\vect{G}',\omega)=\sum_{\vect{G}''}\bigg[\delta_{\vect{G}_1\vect{G}_2}-
\sum_{\vect{G}_3}\chi_s(\vect{q},\vect{G}_1,\vect{G}_3,\omega)\\
\times f_\text{Hxc}(\vect{q},\vect{G}_3,\vect{G}_2,\omega)\bigg]^{-1}_{\vect{G}\vect{G}''}\chi_s(\vect{q},\vect{G}'',\vect{G}',\omega),
\end{multline}
where the xc kernel in reciprocal space was defined in Eq. (\ref{eqn:application:longrange:fxcqspace}).
By calculating $\chi$ on a frequency grid, one thus obtains the optical spectrum (including
a finite broadening in order to make the spectrum smooth).
The size of the matrices involved are determined by the number of $\bfk$-points associated with
the numerical discretization scheme employed.
The spectral contribution from large $\vect{G}$ and $\vect{G}'$ elements in
$\chi$ typically decays rapidly, so only few reciprocal lattice vectors need to be considered.

As discussed in Section \ref{subsec:VI.E}, the head of the xc kernel plays
a dominant role in periodic solids. Fig. \ref{fig:SiSpectrum} shows the experimental
spectrum of Si together with the calculated spectrum of ALDA, which has a vanishing head of the xc matrix.
Besides producing a red-shifted spectrum due to the band-gap problem, the ALDA spectrum lacks
the strong excitonic peak near the gap.
As expected, local and semilocal functionals such as the ALDA break down for the highly non-local excitonic effects.
Big improvements can be achieved by having a finite head in the xc kernel. We now list a few
xc kernels which have the proper long-range behavior that is required for a  finite head of the xc matrix.

The long-range corrected (LRC) kernel \cite{Ghosez1997} is a simple ad-hoc approximation developed mainly for
studying the effect of the long-range behavior. It has the form
\begin{equation}
f_{\rm xc}^{\rm LRC}(\vect{q},\vect{G},\vect{G}',\omega)=-\frac{\alpha}{|\vect{q}+\vect{G}|^2}\: \delta_{\vect{G},\vect{G}'},
\end{equation}
where $\alpha$ is a system-dependent fitting parameter. Despite its simple form, LRC spectra
(with properly chosen $\alpha$) can be in good agreement with
experiments \cite{Sottile2003,Botti2004} since the head contribution
of the kernel usually overwhelms the body contributions (sometimes called local field effects).
A simple connection of the parameter $\alpha$ with the high-frequency dielectric
constant has been suggested \cite{Botti2004}. This xc kernel should not be confused
with the long-range correction in ground-state DFT, where it means a correction
term to fix the rapid decay of local and semilocal xc potentials away from nuclei \cite{Leininger1997}.

The Bethe-Salpeter equation (BSE) \cite{Hanke1980,Onida2002} is a many-body equation for a two-particle
polarization function (which is closely related to the two-particle Green's function) \cite{Stefanucci2013}.
Today, the BSE, combined with the GW method, is the most accurate approach to calculating optical properties of materials.
However, the scaling of the computational cost
versus system size is not favorable; the use of GW-BSE has
therefore been limited to moderate system sizes, despite recent
progress \cite{Puschnig2002,Fuchs2008,Ramos2008,Setten2011}.
From the point of view of TDDFT, the BSE has been an important guide towards the development
of very accurate excitonic xc kernels. The idea is to construct $f_{\rm xc}^{\rm ex}$ via an
integral equation that features the same four-point response functions that are featured in
the BSE \cite{Reining2002,Onida2002}. The resulting xc kernel reproduces the
results of the full BSE \cite{Adragna2003,Marini2003,Sottile2003,Sottile2007,Bruneval2005,Bruneval2006,Gatti2007,Gatti2011}.
However, the computational cost is essentially as high as that of solving the full BSE;
therefore, this xc kernel has mainly served as a proof of concept that TDDFT is capable
of producing accurate excitonic effects. Furthermore, the LRC xc kernel can be shown to
emerge from this BSE-based xc kernel in the long-range limit \cite{Botti2007}.

\begin{figure}
\begin{center}
\includegraphics[angle=0, width=8.6cm]{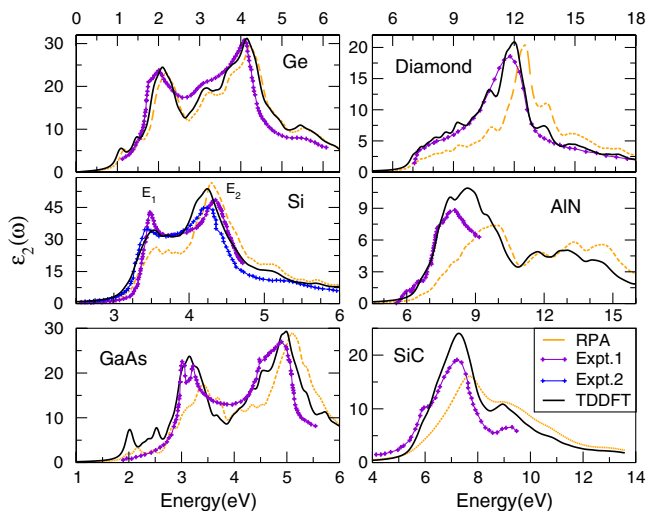}
\end{center}
\caption{Optical absorption spectra of various bulk semiconductors calculated with TDDFT using the bootstrap xc kernel, Eq. (\ref{boot}).
Reproduced with permission from APS from \cite{Sharma2011}. \copyright 2011.}
\label{fig:boot}
\end{figure}

A computationally much simpler alternative is the recently proposed `bootstrap' kernel \cite{Sharma2011,Sharma2012}, defined as
\begin{equation} \label{boot}
f_{\rm xc}^{\rm boot}(\vect{q},\vect{G},\vect{G}',\omega)=\frac{\epsilon^{-1}(\vect{q},\vect{G},\vect{G}',\omega=0)}
{\chi_0(\vect{q},\vect{G}=0,\vect{G}'=0,\omega=0)}.
\end{equation}
Since $\chi_0(\bfq \to 0) \sim q^2$,
the bootstrap kernel has the correct $O(q^{-2})$ long-range behavior.
The bootstrap kernel performs well for a wide range of solids, as illustrated in
Fig. \ref{fig:boot}, and even works for the case of strongly bound excitons such as in solid argon or LiF
(note that the noninteracting response function $\chi_0$ typically contains a band-gap
correction such as a scissor operator or GW).

We also briefly mention that the VS98 meta-GGA \cite{VanVoorhis1998}
has recently shown some promise for calculating optical spectra of insulators with TDDFT \cite{Nazarov2011}.

As an alternative to obtaining optical spectra via the dielectric matrix,
a direct calculation of excitonic binding energies of insulators and semiconductors via
the Casida equation is also possible \cite{Turkowski2009,Yang2012,Yang2013a,Yang2013b}.
The advantage of this approach is that excitonic binding energies---which can be in
the meV range for materials such as GaAs---can be numerically well resolved;
this is much more difficult to do from the dielectric function, which typically yields relatively low-resolution
optical spectra such as in Figs. \ref{fig:SiSpectrum} and \ref{fig:boot}. It is found that the bootstrap kernel
yields good results for strongly bound excitons, but is less accurate for the more weakly bound cases \cite{Yang2013a}.
Accurate triplet exciton binding energies are even more difficult to obtain. The development of xc kernels for excitonic effects
in solids thus remains an important task for future research.

It should be noted that Eqs. \parref{eqn:epsilonmac} and \parref{eqn:inverseepsilon} imply that
the eigenvalues in the Casida equation approach are the poles in $\epsilon^{-1}$ instead of $\epsilon_{\rm mac}$,
so that the absorption peaks are not given directly. This problem is solved through a modification of the Hartree kernel:
\begin{equation}
\bar f_{\rm H}(\bfq,\bfG,\bfG')=\left\{\begin{array}{cl}0 & G=G'=0,\\
\frac{4\pi}{\left|\bfq+\bfG\right|^2}\delta_{\bfG\bfG'} & \text{otherwise}.\end{array}\right.
\end{equation}
By using $\bar f_{\rm H}$ instead of $f_{\rm H}$ in TDDFT, $\epsilon_{\rm mac}$ becomes \cite{Onida2002}
\begin{equation}
\epsilon_{\rm mac}(\omega)=\lim_{\bfq\to0}\left[1-v_{G=0}(\bfq)\bar\chi(\bfq,\bfG=0,\bfG'=0,\omega)\right],
\label{eqn:epsilonmac:noinverse}
\end{equation}
where $\bar\chi$ is the modified response function resulting from TDDFT with $\bar f_{\rm H}$.
Thus the Casida equation with $\bar f_{\rm H}$ yields eigenvalues corresponding to the peaks in the optical spectra.
Since Eq. \parref{eqn:epsilonmac:noinverse} avoids the matrix inversion involved in Eq. \parref{eqn:epsilonmac},
the use of $\bar f_{\rm H}$ is also a standard practice in the response function approach of TDDFT.

\subsubsection{Metallic systems}

The optical properties of metallic systems (bulk metals or metallic nanoparticles) are
strongly determined by the fact that they have a sea of delocalized conduction electrons with a Fermi surface.
Hence, their low-energy elementary excitation are quite different compared to systems with a gap (insulators and semiconductors).
Whereas the outstanding features of the optical spectra of insulators are the {\em excitons}, metallic systems are dominated by
{\em plasmons}.

Excitons and plasmons are observed using different experimental techniques: excitons
are seen in optical absorption spectra (i.e., via coupling to transverse optical fields); on the other hand, plasmons couple to
longitudinal fields, and are thus observed using electron energy loss spectroscopy or inelastic light (or X-ray) scattering
spectroscopy \cite{Quong1993,Balz1997,Huotari2011,Cazzaniga2011}.

\begin{figure}[t]
\begin{center}
\includegraphics[angle=0, width=6.9cm]{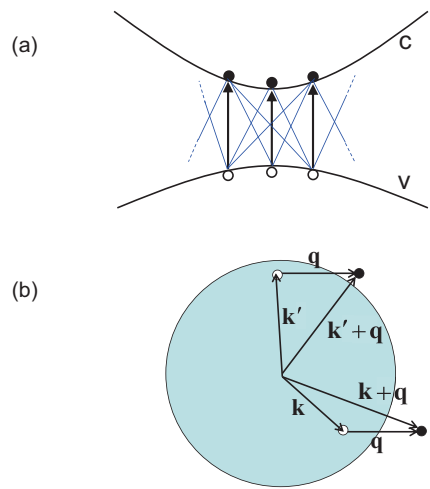}
\end{center}
\caption{(a) Excitons arise from a coupling of single-particle excitations between valence and conduction  band
in an insulator, mediated by dynamical xc effects. (b) Particle-hole excitations with momentum transfer $\bfq$
across the Fermi surface of a simple metal. A plasmon is a coherent superposition of many such excitations, coupled by
Coulomb interactions.
}
\label{fig:coll}
\end{figure}

From a TDDFT perspective, both excitons and plasmons are {\em collective excitations} of the many-body system. However, there is
a big difference as to what causes the collective behavior in the Kohn-Sham system. Excitons can be viewed as a coherent superposition
of a large number of individual particle-hole excitations between valence and conduction band, mediated
via long-range dynamical xc effects \cite{Yang2012b} (see Fig. \ref{fig:coll}). As we discussed in the previous subsection, it is not
easy to find xc kernels which reproduce excitonic effects: all electron-gas based approximations (such as
ALDA) will fail.

\begin{figure}[t]
\begin{center}
\includegraphics[angle=0, width=7.5cm]{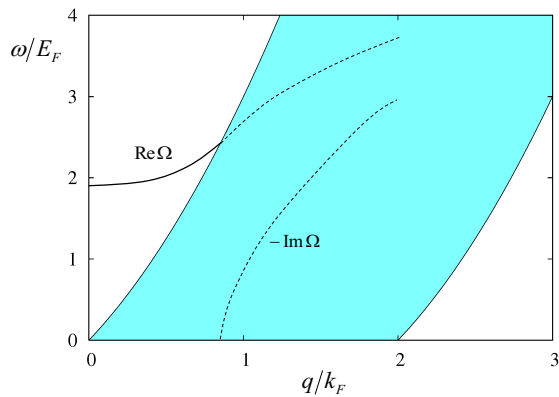}
\end{center}
\caption{Schematic illustration of the particle-hole continuum of a 3D homogeneous electron liquid, and the RPA
plasmon dispersion. In RPA, the plasmon is undamped until it enters the particle-hole continuum, where it
decays into incoherent particle-hole excitations (Landau damping). TDDFT gives very similar results \cite{Tatarczyk2001}.
}
\label{fig:plas1}
\end{figure}

On the other hand, plasmon excitations in metallic systems are relatively easy to capture within TDDFT.
The reason is that plasmons can be viewed as collective charge-density oscillations, and it is a straightforward
textbook exercise in electromagnetism to show that such oscillations arise already from classical electrostatic (RPA-type) interactions;
many-body xc effects only cause relatively minor corrections (but are important and subtle for plasmon damping, see below).
One thus derives the classical plasma frequency as
\begin{equation}
\omega_{\rm pl} = \sqrt{ \frac{4\pi n e^2}{m }} \:.
\end{equation}
The plasmon dispersion of a homogeneous electron liquid can be calculated using TDDFT linear response theory,
along similar lines as finding the zeros of the Lindhard dielectric function \cite{GiulianiVignale}.
The analytic form of the plasmon dispersion up to order $q^2$ is given by
\begin{equation} \label{12.plasmon_lowq}
\Omega(q) = \omega_{\rm pl}\left[1 + \left(\frac{3 k_F^2}{10 \omega_{\rm pl}^2} + \frac{1}{8\pi}f_{\rm xc}(q=0,\omega_{pl})\right)q^2\right],
\end{equation}
where the terms without $f_{\rm xc}$ are the RPA result.
For small $q$, the plasmon lies outside the particle-hole continuum, as illustrated in Fig. \ref{fig:plas1}.
As soon as the plasmon dispersion enters the particle-hole continuum, it becomes subject to Landau damping
(decay into incoherent particle-hole excitations). This damping occurs already in RPA \cite{FetterWalecka}.
But outside the particle-hole continuum, the only source of plasmon damping comes from
the imaginary part of the xc kernel. The physical origin of the
low-$q$ plasmon damping is decay into multiple particle-hole excitations. A frequency-independent $f_{\rm xc}$ (such as the ALDA)
has no imaginary part and hence leaves the plasmon undamped.

\begin{figure}[t]
\begin{center}
\includegraphics[angle=0, width=8.6cm]{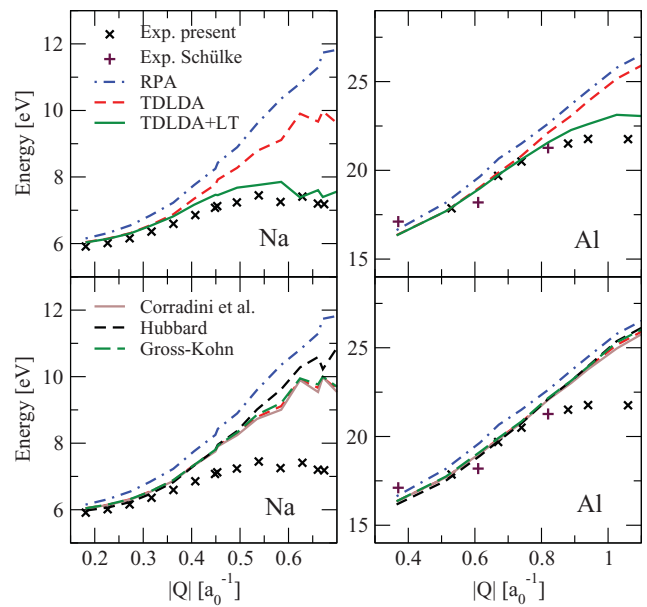}
\end{center}
\caption{Plasmon dispersions of bulk sodium and aluminum: comparison of experiment and TDDFT.
Reproduced with permission from APS from \cite{Cazzaniga2011}. \copyright 2011.)
}
\label{fig:plas2}
\end{figure}

Figure \ref{fig:plas2} shows a comparison of experimental and theoretical results for the plasmon dispersions
of bulk sodium and aluminum \cite{Cazzaniga2011}. The agreement is very good for small plasmon wavevectors,
but for larger wavevectors all TDDFT approaches fail (even the nonadiabatic xc kernel of Gross and Kohn \cite{Gross1985}).
Good agreement is achieved by a hybrid approach in which many-body quasiparticle lifetimes are put by hand into
the response formalism (TDLDA-LT).

Plasmonic effects are found not only in bulk metals, but also in many types of nanostructures.
TDDFT has been extensively used for collective excitations in metallic clusters and nanoparticles.
In general, the results are excellent: plasmon peaks and line shapes for simple metal clusters are very well reproduced, even
at the ALDA level \cite{Calvayrac2000,Morton2010}.
Applications to gold and silver clusters have also been quite successful, and nicely demonstrate
the evolution from atomic-like discrete spectra to plasmon spectra as the cluster size increases
\cite{Aikens2008,Durante2011,Weissker2011}.

A similar picture holds for doped semiconductor nano\-structures such as quantum wells, wires or dots.
Here, collective excitations in the charge and spin channel have been well studied using TDDFT methods;
in general, plasmon dispersions are well reproduced \cite{Ullrich2006}.
The issue of plasmon damping in quantum wells has received a good deal of attention; in particular,
intersubband plasmons in quantum wells have been used to test the Vignale-Kohn approximation of
TDCDFT \cite{Vignale1996,Vignale1997,Qian2003b}, with considerable success \cite{Ullrich1998b,Ullrich2001,Ullrich2002,DAmico2006}.

\section{The future of TDDFT} \label{sec:VIII}

In the final section of our overview, we attempt a forecast of the directions in which the field of TDDFT will be progressing.
We will highlight some areas in which applications of TDDFT are likely to see a lot of activity because of their practical importance.
We will also give a list of issues and challenges---some of them formal, some of them practical---which will
keep the TDDFT community busy for years to come.

{\em Biological Systems.} It has been said that ``if the 20th century was the century of physics, the 21st century will be
the century of biology'' \cite{Venter2004}. Without doubt, DFT and TDDFT methods will play
a key role in the scientific effort to understand the links between structure and functionality
in biochemistry and biology. This is due to the fact that DFT is the only method capable of delivering
ab-initio descriptions of the electronic structure of systems with tens of thousands of atoms;
thanks to the development of linear-scaling methods, even systems with millions of atoms are now within reach
\cite{Bowler2010,Sena2011,Bowler2012}.

Applications of TDDFT for large biomolecules have begun to emerge at a rapid rate
\cite{Varsano2006,Castro2009,Spallanzani2009,Jensen2009,Rozzi2012,Wanko2012,Li2012,Ghane2012}.
Many of these studies are concerned with the electronic and optical properties of DNA fragments,
or the properties of light-harvesting complexes. Apart from the availability of the necessary computer
power (hardware as well as software), there are several developments in DFT which facilitate this trend towards large
organic systems:
\begin{itemize}
\item With the range-separated hybrid functionals, we now have the tools for
describing charge-transfer excitations with TDDFT (see Section \ref{subsec:VI.C}).

\item A new generation of DFT approaches for van-der-Waals interactions has emerged
\cite{Dion2004,Langreth2009,Vydrov2009,Vydrov2010,Dobson2012,Tkatchenko2009,Tkatchenko2012},
which allow for first-principles calculations of the structure of sparse matter,
adsorption on surfaces, and many other applications.

\end{itemize}

{\em Coupled electron-nuclear dynamics.} The coupling of electronic and structural degrees of freedom is a deciding factor in
many functionalities of biological systems. An example are photoinduced processes such as
photoisomerization. As discussed in Section \ref{subsec:VII.C}, TDDFT gives access to excited-state
potential energy surfaces. But things get really interesting when the dynamics goes beyond
the Born-Oppenheimer approximation, giving rise to effects such as structural relaxation
or ultrafast laser-driven molecular reorganization or dissociation. In such situations,
TDDFT can be combined with molecular dynamics, at various levels of sophistication \cite{Marx2009,McEniry2010,Tully2012}.
For a recent review of nonadiabatic dynamics see Ref. \cite{Curchod2013}.

The most straightforward TDDFT approach for coupling electronic and nuclear dynamics is via
the Ehrenfest approximation, which is a mixed quantum-classical treatment where forces on
the classical ions result from a mean-field average over the electronic states.
Ehrenfest dynamics works well in many situations \cite{Castro2004b,Meng2008a,Meng2008b,Alonso2008}, but has its clear limitations for
situations where a branching of ionic trajectories occurs, and where the excited states involve
multiple pathways. Such phenomena can be described with the so-called surface-hopping schemes \cite{Tully1990,Tapavicza2007,Tavernelli2010},
in which multiple excited-state potential energy surfaces can participate in the dynamics, governed by
a stochastic hopping algorithm.

But all of these approaches are based on classical nuclear dynamics and are thus missing out on nuclear quantum effects.
Important effects of nuclear dynamics such as interference, decoherence or tunneling are therefore not captured.
There are already some efforts underway to develop approaches that combine electronic TDDFT with nuclear quantum dynamics
\cite{Kreibich2001,Kreibich2008,Butriy2007,Abedi2010,Abedi2012,Curchod2011a,Curchod2011b}.
It can be expected that the field will continue to advance towards a comprehensive and practical
treatment of electronic and nuclear degrees of freedom. This would open up a large area of interesting
new applications of TDDFT.

{\em Linear and nonlinear optics in materials.} In Section \ref{subsec:VII.D} we discussed how linear-response TDDFT is
applied to describe optical properties of materials (insulators and metals). It can be expected that this will remain
a highly active area of research. Significant progress can be expected along several directions.

There is a need for better xc kernels for solids. It is very likely that these kernels will be expressed in terms of
occupied and unoccupied orbitals, rather than the density. The bootstrap kernel, Eq. (\ref{boot}), is an important
step in the right direction, but it is not so clear how it can be systematically improved.
For instance, a spin-dependent generalization of the bootstrap kernel (which would allow a description
of singlet and triplet excitons) is problematic \cite{Yang2013a}.

A particularly hot area of research are photovoltaic processes in organic systems (polymers or
biological light-harvesting complexes) \cite{Scholes2006,Deibel2010,Bakulin2012,Li2012b}.
There is a rich variety of photophysical processes involved, such as
formation and diffusion of excitons, formation of charge-transfer complexes, relaxation, and charge separation.
At present, no comprehensive ab-initio picture of these processes exists.
This represents one of the major challenges for TDDFT, and should soon be within reach,
based on existing methodologies and new developments. A promising idea is the recently proposed
real-time visualization of exciton dynamics using the time-dependent transition density matrix \cite{Li2011}.

In the majority of applications of TDDFT in periodic solids, the dielectric function (or related response
properties) are calculated, which yield optical spectra or scattering cross sections. But there are many
nonlinear or explicitly time-dependent processes of interest, which go beyond response theory and require, in principle,
a time-dependent calculation. Real-time TDDFT calculations for periodic solids are beginning to emerge
\cite{Otobe2008,Shinohara2010,Yabana2012,Shinohara2012} to simulate hot carrier generation, dielectric breakdown, and coherent phonons
in semiconductors and insulators. Such calculations, in particular if light propagation effects are included via a coupling
with Maxwell's equation, pose a significant computational challenge and call for the development of new multiscale
or multidomain approaches \cite{Varga2010,Goncharov2011}.

{\em Other developments.} Let us conclude with a mixed bag of various formal and practical challenges and unsolved problems
for present and future TDDFT research.

\begin{itemize}

\item {\em Nonadiabatic xc functionals.} Nonadiabatic xc functionals are needed
for double excitations in finite systems, for dissipation in extended systems, for
exciton Rydberg series, for conical intersections, and many other important phenomena.
Electron-gas based functionals \cite{Vignale1996,Vignale1997} are of limited use \cite{Ullrich2006b}; a connection
with many-body approaches seems the most promising avenue towards the development of simple, practically
useful nonadiabatic functionals \cite{Gritsenko2009,Romaniello2009,Sangalli2011,Sakkinen2012}.
Another possibility could be via reduced  density-matrix-functional theory
\cite{Giesbertz2008,Rajam2010,Pernal2012,Elliott2012}.

\item {\em Open systems.} TDDFT for open systems is of interest for the description of
transport through nano- or mesoscopic systems, where a region of interest (e.g, a molecule) is connected to
energy and particle reservoirs via metallic leads \cite{DiVentra}. It is also of interest
for treating dissipative dynamics. The coupling to a reservoir can be formally
treated within TDDFT in various ways: with a master equation approach \cite{Burke2005a}, using stochastic methods
\cite{DiVentra2007,DAgosta2008,Biele2012},
and by mapping the open physical system onto a noninteracting closed system \cite{YuenZhou2009,YuenZhou2010,Tempel2011}.
The formal aspects are complicated and subject of ongoing debate \cite{DAgosta2013}; practical xc functionals
for open systems and applications beyond simple model systems can be expected in the future.

\item {\em Strongly correlated systems.} There has been some interesting recent work in which
TDDFT methods were successfully applied to the transport in strongly correlated model lattice systems exhibiting Coulomb blockade and  the
Kondo effect \cite{Verdozzi2008,Kurth2010,Stefanucci2011,Liu2012}. A subtle feature of the xc potential, its derivative discontinuity
upon change of particle number (briefly mentioned in Section \ref{subsec:VII.D}), turns out to be crucial for capturing these
effects. Most of these studies are for one-dimensional Hubbard-type lattice systems \cite{Lima2003,Capelle2013}, but three-dimensional systems
were also considered \cite{Karlsson2011,Verdozzi2011}. In the future, work along these lines is likely to
make an impact in the description of realistic strongly correlated systems and materials, which so far have remained problematic for (TD)DFT.

\item {\em Extensions of the formalism.} Ground-state DFT has long ago been extended to
finite temperatures \cite{Mermin1965} and to relativistic systems \cite{Engel2011}. The corresponding TDDFT versions
are not yet available, but would be of great interest for matter under extreme conditions.
Finite-temperature TDDFT, which might include elements of nonequilibrium thermodynamics and time-dependent thermal ensembles, could also be of
interest for thermal transport and thermoelectric properties. Relativistic TDDFT has been used for calculating
molecular excitation energies and response properties \cite{Toffoli2002,Gao2004,Gao2005,Salek2005,Henriksson2008}, and real-time Dirac-Kohn-Sham
calculations have been explored \cite{Belpassi2011}, but formally rigorous general existence proofs have yet to be worked out.
Some promising developments have recently occurred in the application of TDDFT methods for quantum electrodynamics
\cite{Ruggenthaler_qed2011,Tokatly2013}.

\end{itemize}

\acknowledgments

This paper is based in part on a two-weeks course on excitations in materials, delivered in July 2012 at
the Universidade Federal do ABC at Santo Andr\'e, sponsored by the American Physical Society
and the Sociedade Brasileira de F\'isica. We also acknowledge support from NSF Grant No. DMR-1005651.
We thank Yonghui Li for preparing Figures \ref{figCO2} and \ref{figCO2_strong}.

\bibliography{TDDFT_review}

\end{document}